\documentstyle[11pt,psfig]{l-aa}
\begin{document}
\thesaurus{
               09. 
              (02.08.1;  
               02.19.1;  
               02.23.1;  
               06.03.1;  
               06.15.1)  
                            }
\title{Acoustic wave propagation in the solar atmosphere}
\subtitle{V.~Observations versus simulations}
\author{ J.~Theurer \inst{1}, P.~Ulmschneider \inst{1} 
and W.~Kalkofen \inst{2}}
\institute{  Institut f\"ur Theoretische
             Astrophysik der Universit\"at
             Heidelberg, Tiergartenstr.~15,\newline
             D--69121 Heidelberg, Germany
             \and 
             Harvard-Smithsonian Center for Astrophysics, 
             60 Garden St.,\newline
             Cambridge, MA 02138, USA
} 
\offprints{P.~Ulmschneider}
\date{Received 24.~Jan.~1997; accepted 27.~Mar.~1997}
\maketitle
\begin{abstract}
We study the evolution of spectra of \hbox{acoustic} waves
that are generated in the convection zone 
and propagate upward into the photosphere, where we
compare the simulated acoustic spectra with 
the spectrum observed in an Fe I line. 
Although there is no pronounced 3~min component
in the spectra generated in the convection zone,
there are dominant 3~min features in the theoretical
spectra, in agreement with the observed spectrum.
We interpret the occurrence of the 3~min features as
the response of the solar atmosphere to the acoustic waves
which shifts high frequency wave energy to low frequencies.
We also find qualitative agreement for the acoustic power
between the wave simulations and the observations.
\end{abstract}
\section {Introduction}
In the present series of papers we investigate the response of 
the solar atmosphere to acoustic wave excitations. In 
Paper I (Sutmann \& Ulmschneider 1995a) we studied the 
excitation and propagation of small-amplitude (linear) adiabatic 
waves, and in  Paper II (Sutmann \& Ulmschneider 1995b),
the excitation of large-amplitude adiabatic waves which 
form shocks. Similar numerical and analytical work was 
carried out by Kalkofen et al.~(1994). These 
studies show that, for monochromatic or 
pulse excitation, the solar atmosphere 
responds with oscillations in an added band of frequencies
generally above $5.5$~mHz (period $\sim 3$~min) and that,
depending on the type of excitation, these 3~min oscillations 
usually decay after some time. Because both the monochromatic 
and pulse excitation happen to an atmosphere considered to be 
initially at rest, these 3~min oscillations could be
seen as decaying transients. Yet Paper I 
also showed that even with very small amplitude (linear) 
excitations, for the stochastic wave case, the generated 3~min type 
oscillations did not decay as in a switch-on effect, but were 
perpetually regenerated due to the stochastic nature of the 
excitation (cf.~Fig.~14 of Paper I). In Paper II, in addition to 
monochromatic or pulse excitations, perturbations by four 
different types (Gaussian, box, exponential and stochastic) of 
acoustic wave spectra were studied. It was found that these 
perturbations led to 3~min oscillations which over the time span 
of the calculations ($\sim 2500$~s) did not decay.
Below we will refer to the 3~min oscillations near the acoustic cutoff as
``resonance oscillations''.

The 3~min oscillations are also a prominent 
observational signal in the outer solar atmosphere. 
In the quiet solar atmosphere they are seen 
in the interior of supergranulation cells: in the Ca 
II H and K lines, in H$\alpha$ and in the Ca II infrared triplet lines 
as well as in other, lower-lying 
photospheric and chromospheric lines. 
For detailed reviews and discussions of the 3~min oscillations 
see Deubner
(1991), Fleck \& Schmitz (1991), Rutten \& Uitenbroek (1991), 
Rossi et al.~(1992), Carlsson \& Stein 
(1994), Rutten (1995, 1996), Steffens et al.~(1995) and Al (1996). 
 
The analytical work of Paper I will be extended to include
the excitation of linear, 
adiabatic waves by a pulse and by acoustic spectra 
(Sutmann et al.~1997; Paper III). For similar analytical
work see Schmitz \& Fleck (1995).
Recently (Theurer et al.~1997; Paper IV) we investigated the excitation of
the solar atmosphere by large-amplitude, non-adiabatic acoustic 
spectra. In Paper II the excitations with acoustic spectra could 
not be carried out over extended times because due to shock 
formation and shock heating the adiabatic waves generate secular 
changes of the atmospheric structure with temperatures which 
perpetually rise with time. The non-adiabatic treatment in Paper~IV
avoids this problem. Here the excitation leads to a dynamical 
equilibrium between shock heating and radiative cooling, which 
results in a time-independent mean atmosphere, where the wave 
excitation could now be carried out up to arbitrarily large 
times. 

As already suspected in 
Paper II these calculations showed that the 3~min type resonance oscillations 
near the acoustic cutoff now remain a permanent, non-decaying feature of the 
wave spectrum and that these 3~min ``resonances''   
became more and more prominent with height.
It was found that 
regardless of the initial shape of the acoustic wave spectrum 
(Gaussian or stochastic), introduced at 
$z = 0$~km into the atmosphere, the resulting spectrum 
at a height of $z = 2000$~km consisted almost exclusively of 
components in the 3~min band. Moreover, after about 500~s the 
spectra at a given height no longer showed any time dependence. 

In an attempt to simulate the observed 3~min 
oscillations, in particular, those observed in
the K$_{2v}$ bright points in the interior of supergranulation 
cells, Carlsson \& Stein (1994, 1995) excited the 
oscillations using the observed velocity variation of 
a low-lying Fe I line (Lites et al.~1993).
They obtained good agreement of observed
and simulated H line profiles as well as of 
the time delay between the oscillations at the Fe line
and the H line.

Similar work by Cheng \& Yi (1996) 
but employing a different numerical code has recently been 
carried out. They introduced high-frequency 
acoustic wave power in addition to the observed spectrum and 
found that high-frequency waves did not contribute to the 
3~min band at chromospheric heights,
concluding that the pronounced power in the 3~min band must 
already be 
present at the site of generation of
the waves. This claim has to be viewed with some
caution since their code, which is
different from that of Carlsson \& Stein, does not have 
an adaptive mesh capability and thus does not allow 
the treatment of shocks and hence 
the limiting shock strength behaviour and the merging
of shocks, that is, of effects we find critical for
the generation of power near the acoustic cutoff
from high-frequency waves. 
However, the question remains open whether the 3~min 
oscillations are introduced into the acoustic spectrum 
at the height where the waves are 
generated, or added later by the atmosphere. 

In recent years it has become clear that the acoustic energy 
generation in the convection zones of late-type stars is 
strongly tied to the Kolmogorov turbulent energy spectrum, 
which is now well supported 
for the solar convection zone by numerical  
simulations (Cattaneo et al.~1991) as well as by high-resolution 
observations (Muller 1989, Nesis et al.~1993). For a discussion 
see Musielak et al.~(1994). On the 
basis of this turbulent energy spectrum, the wave generation 
calculations using the Lighthill-Stein theory of sound 
generation produce acoustic spectra that are smooth 
and have a single broad peak in the 1~min band, i.e., most of 
the acoustic power is found at frequencies much higher than the 
3~min band (Figs.~3 and 4). This is due to the property of 
the Kolmogorov turbulent cascade in its inertial range 
(discussed below) of extending to frequencies of at least 
100 Hz under adiabatic conditions when viscosity limits the 
cascade, or to frequencies of at least 300 mHz when radiative 
exchange between adjacent bubbles limits the cascade. 

Thus one must reconcile the theory showing the 
generation of a high-frequency wave spectrum without a noticeable 3~min
feature with the observations which show a low-frequency 
spectrum dominated by the 3~min band. The aim of the present 
work is to find an answer to this problem. A possible answer is 
suggested by the results of the previous papers of this series 
and by the work of Kalkofen et al.~(1994)  showing that
3~min oscillations can be the 
response of the atmosphere to wave excitations. We will argue that 
the 3~min component is a feature that is added by the 
atmosphere to the acoustic spectrum by shifting 
power from high frequencies to low frequencies
during the propagation of the waves
from the convection zone to the photosphere. 

In our previous work we found two ways by which such a shift is
accomplished. Fig.~14 of Paper I shows that already for a wave
excitation with very small amplitude, resulting from an acoustic
spectrum with a stochastically changing wave period, one obtains
3~min type resonance oscillations which continuously get
regenerated. The second way, as shown in Paper IV, is that
acoustic wave spectra with larger amplitudes lead to
shock formation and shock merging, which generates resonance
oscillations. In both cases the stochastic nature of the
acoustic wave spectrum leads to oscillations which persist in time.

Note that in our computations it
is necessary to distinguish between the quiet 
interior of supergranulation cells and the K$_{2v}$ bright points
that are embedded in it. The work reported here concerns mainly 
the oscillations in the quiet background of 
the cell interior and not 
those in the K$_{2v}$ bright points (cf.~Kalkofen 1996), where 
the periods tend to be somewhat longer than those found 
in this investigation.

In order to test the hypothesis of the transfer of power from
high to low frequencies we have made a number of 
simplifications, such as assuming adiabatic conditions for the 
atmospheric wave calculations and neglecting effects of the 
Earth's atmosphere as well as instrumental effects when 
comparing our results with the observations. These 
simplifications will be removed in subsequent work.  Section 2 
outlines the numerical methods, Sect.~3 presents the results 
and Sect.~4 gives the conclusions. 
 
\section{Method}
\subsection{Initial wave spectrum}
For the treatment of the acoustic wave spectrum 
we follow Paper II. At the lower boundary of the 
computational domain, 
velocity fluctuations are prescribed for the piston. The 
frequency spectrum has $N+1=101$ partial waves at the frequencies 
$\omega_n = 2\pi \nu_n$, 
which are spaced 0.5~mHz apart; and
$\nu_{1} = 4.0$~mHz. 
The piston velocity, $v$, is prescribed by the superposition 
 $$
v(t) = \sum_{n=0}^N u_n \  \sin (\omega_n t + \varphi_n)
\ \ \ ,\eqno(1)$$ 
where $u_n$ is the velocity
amplitude of a partial wave and $\varphi_n$ an arbitrary
but constant phase angle. Since the velocity amplitude at the
piston is small the mechanical energy flux, $F_{\rm A}$, 
can be written using linear theory as
  $$ 
F_{\rm A} =  {1 \over T} \int_0^T \rho c  \sum_{n=0}^N  
                    \left(u_n  \sin (\omega_n t + \varphi_n)\right)^2
                      \cos\alpha_n dt ,\eqno(2)$$
where $T$ is an arbitrary but sufficiently long time for the 
averaging, $\rho$ is the 
density and $\alpha_n$ the phase shift between velocity and 
pressure fluctuations, which is given by 
  $$
\alpha_n = \arctan \left( {\omega_{\rm A}  \over 
\sqrt{ \omega_n^2-\omega_{\rm A}^2} } \right)
\ \ \ ,\eqno(3)$$
where $\omega_{\rm A}=\gamma g/(2 c)$ is the acoustic cut-off 
frequency at the bottom of the atmosphere. We normalize 
the velocity amplitudes to the prescribed acoustic energy 
flux $F_{\rm A}$ and obtain
$$
u_n = q\ \sqrt{ {F_{\rm A} \over {\rho c}}} \ f(\omega_n)
\ \ \ ,\eqno(4)$$
where $q$ is a normalization factor and $f(\omega_n)$ the 
spectral distribution function. The normalization factor $q$ is 
chosen such that upon introduction of Eqs.~(3) and (4) into 
Eq.~(2) 
the prescribed flux $F_{\rm A}$ is obtained. 

\subsection{Kolmogorov turbulent cascade, high frequencies}
The initial acoustic wave spectrum is not well known since it 
results from the turbulent energy spectrum of the solar 
convection, which is poorly known. Musielak et 
al.~(1994) argued on the basis of observations and numerical 
convection simulations, that the turbulence in the solar 
convection zone must have a Kolmogorov type energy spectrum 
in which the wave numbers near $k_0=2\pi /H$, where $H$ is the local 
scale height, represent the energy-containing eddies that start 
the turbulent cascade. 

Since it has been suggested frequently on the basis of observations 
that high-frequency acoustic waves not only do not have any 
influence on chromospheric oscillations but do not even 
exist, we briefly discuss here the high-frequency part of the 
turbulent energy spectrum, which is the origin of the
high-frequency part of the acoustic wave spectrum. 

In homogeneous isotropic 
turbulence, the inertial forces break up larger 
turbulent eddies into smaller ones. A turbulent flow field can 
be described by three characteristic quantities, 
namely, the density $\rho$, 
the bubble size $l_k=2\pi/k$, and the mean velocity $u_k$ of such
bubbles, where $k$ is the wavenumber. It is easily seen that from 
these three quantities only one combination can be formed  
for the heating rate (in erg~cm$^{-3}$~s$^{-1}$), namely: 
  $$ 
\Phi_k=\rho{u_k^3\over l_k}
\ \ \ .\eqno(5)$$
Since in the turbulent cascade the energy is passed on from
large to small bubbles the heating rates must
satisfy the condition 
$\Phi_{k0}=\Phi_{k1}=\Phi_{k2}= \ldots 
={\rm const}$, 
where the wave numbers, 
$k0$, $k1$, $k2, \ldots $, represent a 
series of bubbles of decreasing size. 
As the mean density $\rho$ is unaffected this implies that 
$$ 
u_k\sim l_k^{1/3} 
\ \ .\eqno(6)$$
Equation (6) is referred to as 
the {\it Kolmogorov law}. The range $l_{k0}\ldots l_{kn}$ of 
validity of this law is called the {\it inertial range}. It 
ends when the bubble size $l_k$ is so small that 
viscous heating becomes important, that is, when the turbulent 
heating rate becomes equal to the viscous heating rate, i.e.,
$$ 
\eta \left({{\rm d} u\over {\rm d}z}\right)^2 =\rho{u_{k0}^3\over l_{k0}}
\ \ \ .\eqno(7)$$
For the energy-carrying eddies at the beginning of the cascade we 
have $u_{k0}\approx 1\cdot 10^5$~cm/s, $l_{k0}\approx 1.5\cdot 
10^7$~cm, $\rho\approx 3\cdot 10^{-7}$~g/cm$^3$. With 
$\left({{\rm d} u/ {\rm d}z}\right)^2= (u_{k}/l_{k})^2$ and Eq.~(6) one 
finds for the size of the smallest bubbles
$$
l_k=\left({\eta\ l_{k0}^{1/3}\over \rho u_{k0}}\right)^{3/4} 
\ \ \ .\eqno(8)$$
Using $\eta\approx 5\cdot 10^{-4}$~dyn~s/cm$^2$ we obtain
$l_k= 2.9$~cm and $u_k=290$~cm/s for the smallest bubbles, 
which gives $\nu_k=u_k/l_k=99$~Hz for the maximal frequency. 

This limit of the turbulent cascade is valid only 
under adiabatic conditions. If radiation is considered, 
neighboring bubbles will exchange energy. If 
the smallest size of the bubbles is limited by radiative exchange, it 
will be given by the optical diameter $\tau=l_k\kappa\rho =0.1$. 
Using the analytic representation $\kappa=1.38\cdot10^{-
23} p^{0.738} T^5$~cm$^2$/g for the Rosseland grey opacity 
(cf.~Ulmschneider et al.~1978) and the 
values $T=8300$~K and $p=2\cdot 10^5$~dyn/cm$^2$ for the 
temperature and pressure, we obtain $l_k= 7.5\cdot 
10^4$~cm and $u_k=1.7\cdot 10^4$~cm/s for the smallest bubbles 
with a frequency of $\nu_k=230$~mHz. 
This shows that the acoustic spectrum
extends to relatively high frequencies at 
the point of its generation and confirms that short period 
acoustic waves are indeed present in stellar atmospheres.

\subsection{Kolmogorov turbulent cascade, low frequencies}
While the high-frequency part of the turbulent energy 
spectrum appears to be reasonably well established, there 
is considerable uncertainty about the low-frequency part. 
Musielak et al.~(1994) pointed out that, at low wave numbers, 
observations e.g.~by Muller (1989) show that the velocity power 
still increases with decreasing wave number $k$ although with a 
smaller slope. To take into account a deviation from the 
classical Kolmogorov spectrum for wave numbers $k<k_0$,  
Musielak et al.~considered three possible forms of the energy spectrum, 
which for $k>k_0$ has the Kolmogorov shape but for lower wave 
numbers has different dependences on $k$. Their spectra, called 
extended, broadened and raised Kolmogorov spectra, are given by 
their Eqs.~(35) to (37) and are displayed in their Fig.~1. 

Since we cannot ignore the uncertainty in the turbulent energy 
spectrum we selected the two most extreme cases studied by Musielak et 
al.~(1994) and computed the acoustic energy generation on the basis 
of the extended (eKmG) and of the raised Kolmogorov (rKmG) spectrum, 
both with a modified Gaussian frequency factor. The spectral 
shape of the resulting acoustic spectra, $f(\omega)$, is shown in Figs.~4 
and 8 of Musielak et al.~(1994). Taking a mixing-length parameter of 
$\alpha=2$ we have a total acoustic flux of $F_{\rm A}=1.97\cdot 10^8$ 
and $8.75\cdot 10^7$~erg~cm$^{-2}$~s$^{-1}$ for the eKmG and rKmG 
spectra, respectively. Spline-fitted versions of the selected 
spectra $f(\omega)$ are shown in Figs.~3 and 4 below. 

These spectra are generated in a very narrow height range around 
$z=-130$~km in the convection zone, where the convective
velocity $u$ has its maximum (see Fig.~7 of Musielak et al.~1994).
The reason for the localization of the acoustic energy
generation in a thin layer is the sensitive $u^8$ 
dependence of the quadrupole
sound generation rate on the convective velocity $u$.
Such a height ($z\approx -140$~km) of the maximum acoustic
energy production has also been found by Kumar (1994). Moreover,
it is found (Steffen 1992) that a significant difference of the
mixing-length approach as compared with the numerical convection
computations is that in the latter, the maximum convective
velocity occurs about 100~km deeper, which translates into a
layer of maximum acoustic energy generation, roughly in the
range $z=-150$~km to $-250$~km. We thus feel that a value $z=-160$~km
is a reasonable choice.
\begin{figure*}[t] 
\begin {minipage}[t]{0.485 \textwidth}
\psfig{figure=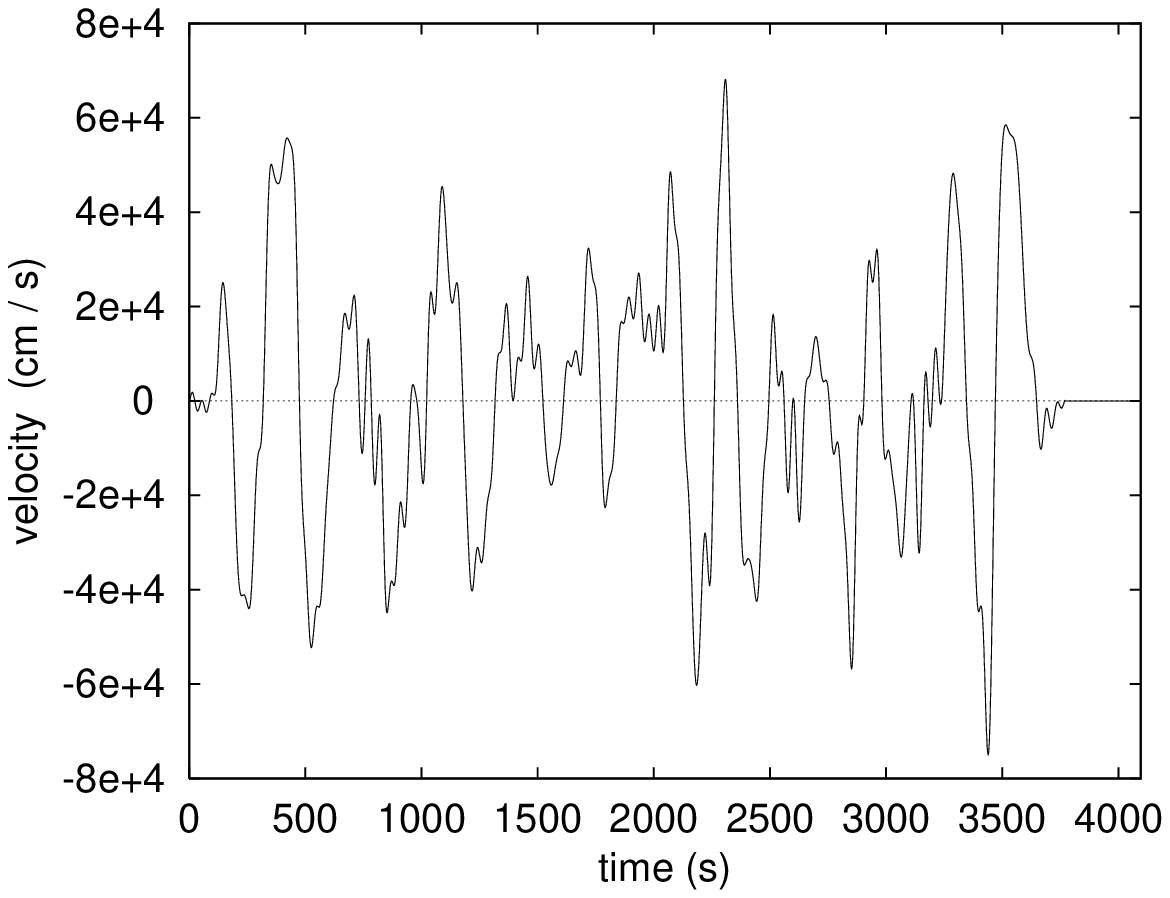,height=6.00cm,width=1.0 \textwidth}
\end{minipage}
\hfill
\begin {minipage}[t]{0.485 \textwidth}
\psfig{figure=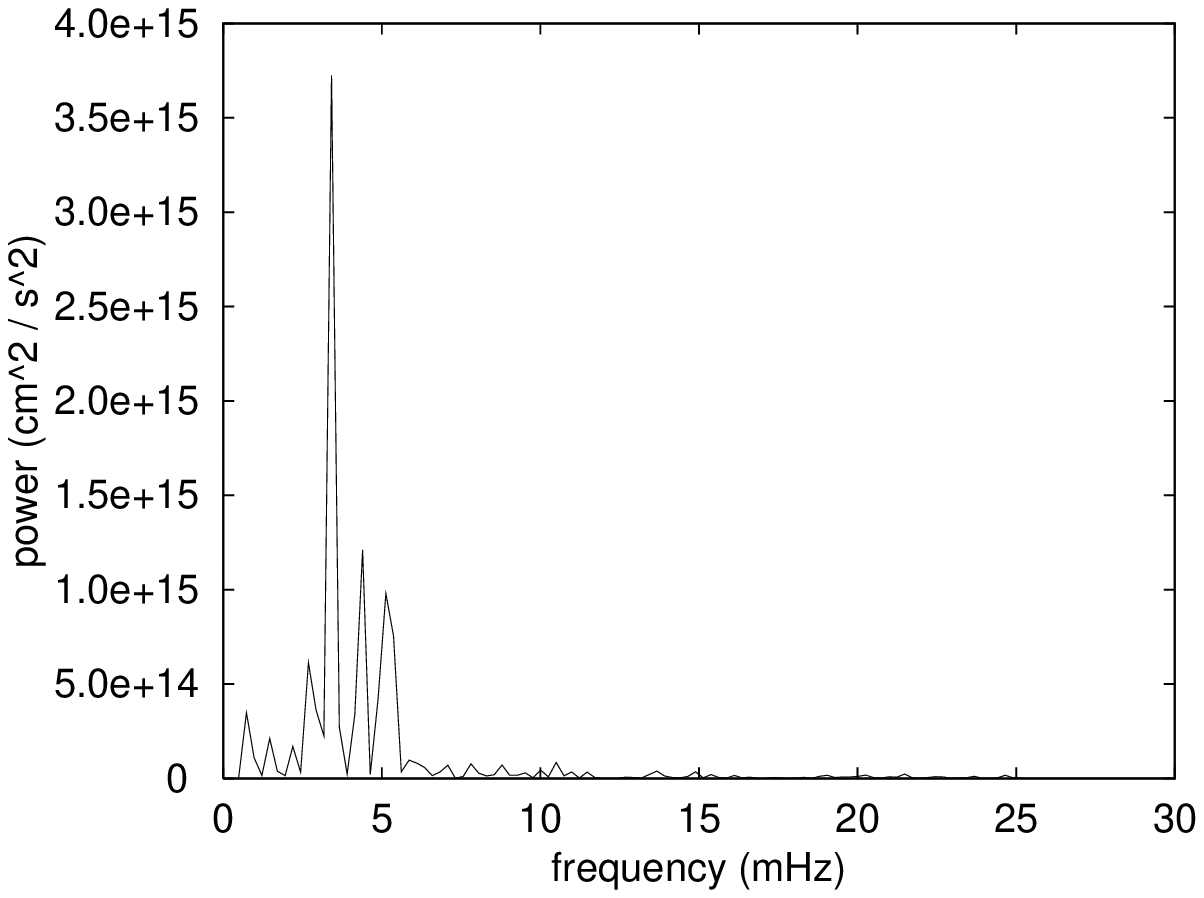,height=6.00cm,width=1.0 \textwidth}
\end{minipage}
\caption{
Velocity (left panel) and power spectrum (right panel) in an Fe I
line observed by Lites et al.~(1993). Some high frequency 
components have been filtered out, the line is formed roughly at 
a height of $z = 250\ {\rm km}$.
} 
\end {figure*} 

As to the choice of the mixing-length parameter $\alpha$, we 
point out that currently the larger values of $\alpha= 1.7$ to $2.0$ 
are preferred (Kumar 1994, Musielak et al.~1994) and that a value 
of $\alpha=2.0$ is indicated by a comparison of the peak value 
of the convective velocity in mixing-length calculations with 
that in time-dependent hydrodynamic 
convection simulations (Steffen 1992), as well as by a careful fitting of 
evolutionary tracks of the Sun 
with its present properties $L$, $T_{\rm eff}$ and age (H\"unsch \& 
Schr\"oder 1996, Schr\"oder \& Eggleton 1996). 
 
\subsection{Wave calculation, atmospheric model} 
We follow the development of the acoustic wave spectrum with a 
hydrodynamic code where for simplicity we assume adiabatic 
conditions. The details of the wave computation have been 
described in Papers I and II. The time-dependent hydrodynamic 
equations are solved using the method of characteristics. We 
follow the development of the originally linear wave, introduced by 
specifying the piston velocity using Eq.~(1) at the bottom of the 
atmosphere, up to the point of shock formation and beyond.  
Shocks are treated as discontinuities and are allowed to grow to 
arbitrary strength and to merge.

We consider an atmospheric slab extending from $z = -
160$~km to a height of 1700~km and use a total of 300 grid 
points with a spacing of 6.2~km. In addition to the fixed 
number of regular grid points there is an arbitrary number of 
shock points, which are allowed to move between the regular 
points according to the speed of the shocks.  

For the atmospheric model we take model C of Vernazza et al.~(1981).
Since for both mixing-length models and 
numerical convection zone 
models the maximum of the convective velocities 
occurs below their lowest point, at $z= - 75$~km, 
we extended this model for a few points 
by fitting a solar convection zone model similar to those 
described by Ulmschneider et al.~(1996).  

\subsection{The observed frequency spectrum} 
For the observed velocity spectrum we use 
the velocity fluctuations of an 
Fe I line formed at $z = 250$~km observed by Lites et al.~(1993). 
The original data consist of a 
time series of Doppler velocities at 5 adjacent  
spatial points along the slit position through
the interior of a supergranulation cell.
The data were low-pass filtered, 
at 15-25~mHz to remove noise, and apodized. They  
were provided to us by Mats Carlsson (1995, private 
communication). Each time series consists of 754 measurements at 
equal time steps of 5~s. This 
velocity spectrum is shown in Fig.~1, with
the velocity fluctuations in the left panel and
Fourier spectrum in the right panel. 
To obtain a set of data points consisting of a number of points
that is equal to a power of 2, for the 
Fast Fourier Transform, we have augmented the data set with zeros up 
to 4096~s, as indicated in Fig.~1. Similar although noisier 
data of a Na I line, also by Lites et al.~(1993), were used by 
Cheng \& Yi (1996, their Fig.~2).  

The dominant feature in Fig.~1 is the well known solar 5~min 
oscillation at 3.3~mHz. This feature is due to trapped 
subphotospheric acoustic eigenmodes involving the entire Sun and 
thus cannot be expected to be reproduced by our local 
one-dimensional simulation. Our focus of interest rather is the 
3~min oscillations in the 5 to 6~mHz range which we want to 
explain as the result of atmospheric resonance oscillations. 
 
For the comparison of our simulations with solar data we note also
the properties of the observed velocity power spectrum at H$_3$ in
the core of the H line by Lites et al.~(1993, their Fig.~6), 
which shows peaks at 5~mHz, 6.5~mHz and 7~mHz. These observations
are from the interior of supergranulation cells and thus both the
quiet background and the K$_{2v}$ bright points contribute
to the spectrum. For the K$_{2v}$
bright points, the observations of v.~Uexk\"ull \& Kneer (1995)
indicate that their emission is spatially and temporally intermittent.
This property is reproduced by a model with impulsive excitation of
waves (Kalkofen 1996). Excitation by the turbulence of the convection
zone cannot reproduce the observed intermittence. The work
reported here therefore concerns waves in the quiet 
cell interior. Nevertheless, the
turbulence-generated waves may also contribute to some bright 
point heating. 
 
\begin{figure}[t]
\begin {minipage}[t]{0.485 \textwidth}
\psfig{figure=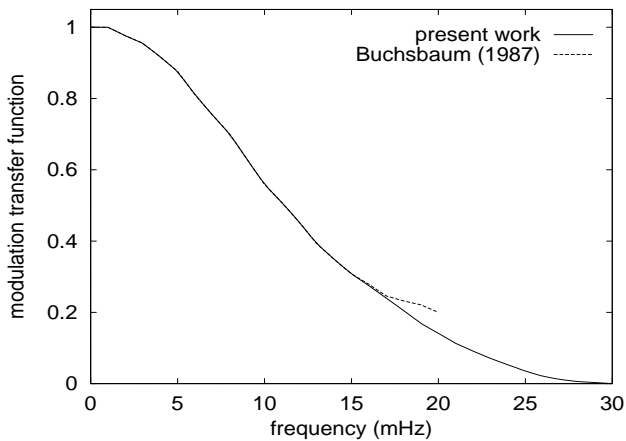,height=6.00cm,width=1.0 \textwidth}
\end{minipage}
\caption{Modulation transfer function after Buchsbaum (1987) for 
the Fe 5930 \protect{\AA} line.
} 
\end{figure} 
\subsection{The modulation transfer function} 
A direct comparison of observed velocity spectra with 
theoretical spectra is not easily achieved since any velocity 
event occurring on the Sun is affected by seeing 
because of the Earth's atmosphere, by instrumental 
degradation and, most of all, by the fact that the observed 
spectral lines have a width $\Delta z_{\rm W}$ of their contribution 
function which extends over at least 
two scale heights. It is 
well known that velocity fluctuations with wavelengths $\lambda 
\ll \Delta z_{\rm W}$ do not contribute to Doppler shifts of the line 
core but instead to line-broadening. 
Thus, when the frequency approaches the critical value where 
$\lambda \approx \Delta z_{\rm W}$, 
the velocity fluctuations seen by 
Doppler shifts abruptly decrease and approach zero for 
higher frequencies. The ratio of the actual velocity to the 
velocity seen as a Doppler shift, $v/v_{\rm Doppler}$, 
which is a function of the acoustic frequency, 
is called the modulation transfer function (see also
Deubner et al.~1988, Ulmschneider 1990).
It is obtained by computing 
propagating sound waves, simulating the spectral lines and 
comparing the Doppler shifts of the absorption core of the line 
with the actual velocities in the line-forming region. 

In this exploratory investigation we avoid such 
detailed line simulations and take published values of the 
modulation transfer function. Figure~2 shows a typical case
from Buchsbaum (1987) for the Fe 5930 line, which 
is formed at a height of  $z=250$~km. The function has 
been extended arbitrarily to 30~mHz and assumed 
to be zero above this frequency. 
\begin{figure*}[t] 
\begin {minipage}[t]{0.33 \textwidth}
\psfig{figure=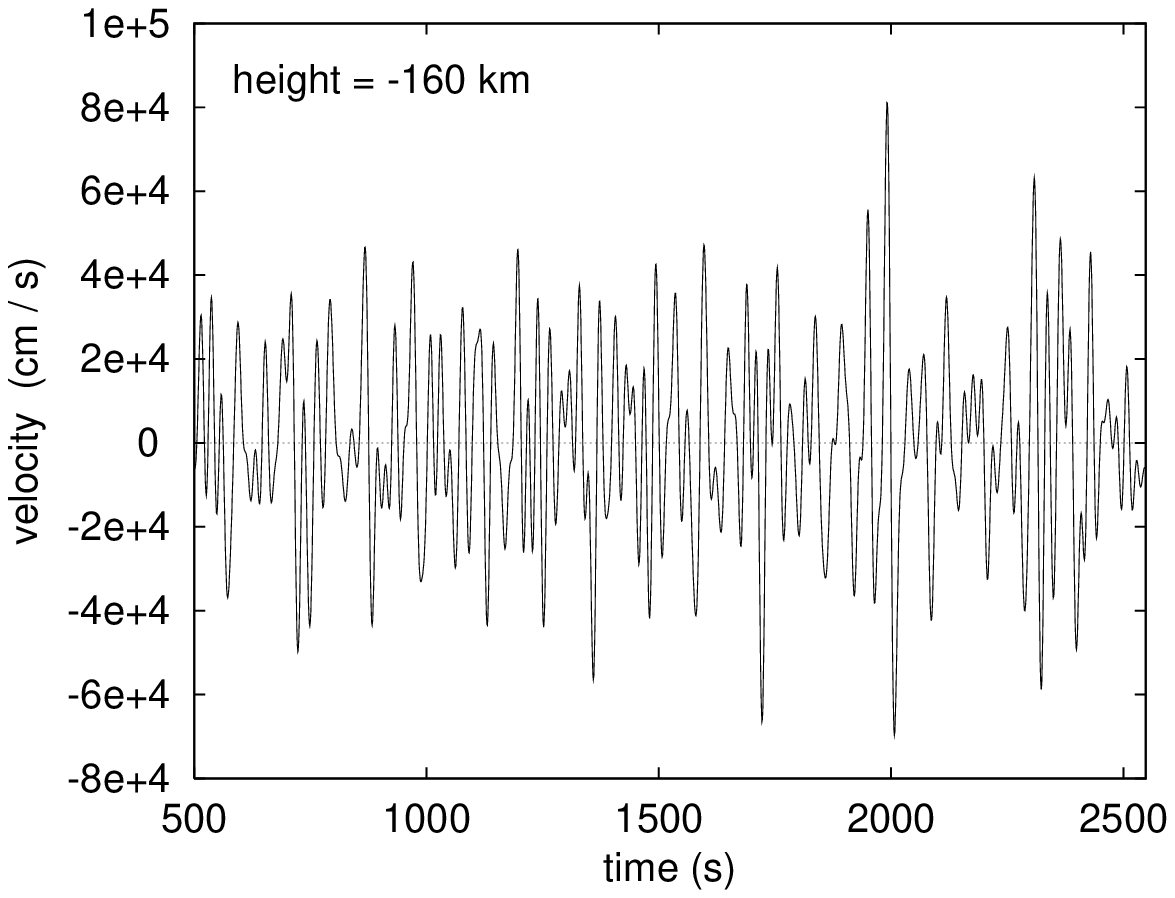,height=3.980cm,width=1.0 \textwidth}
\end{minipage}
\begin {minipage}[t]{0.33 \textwidth}
\psfig{figure=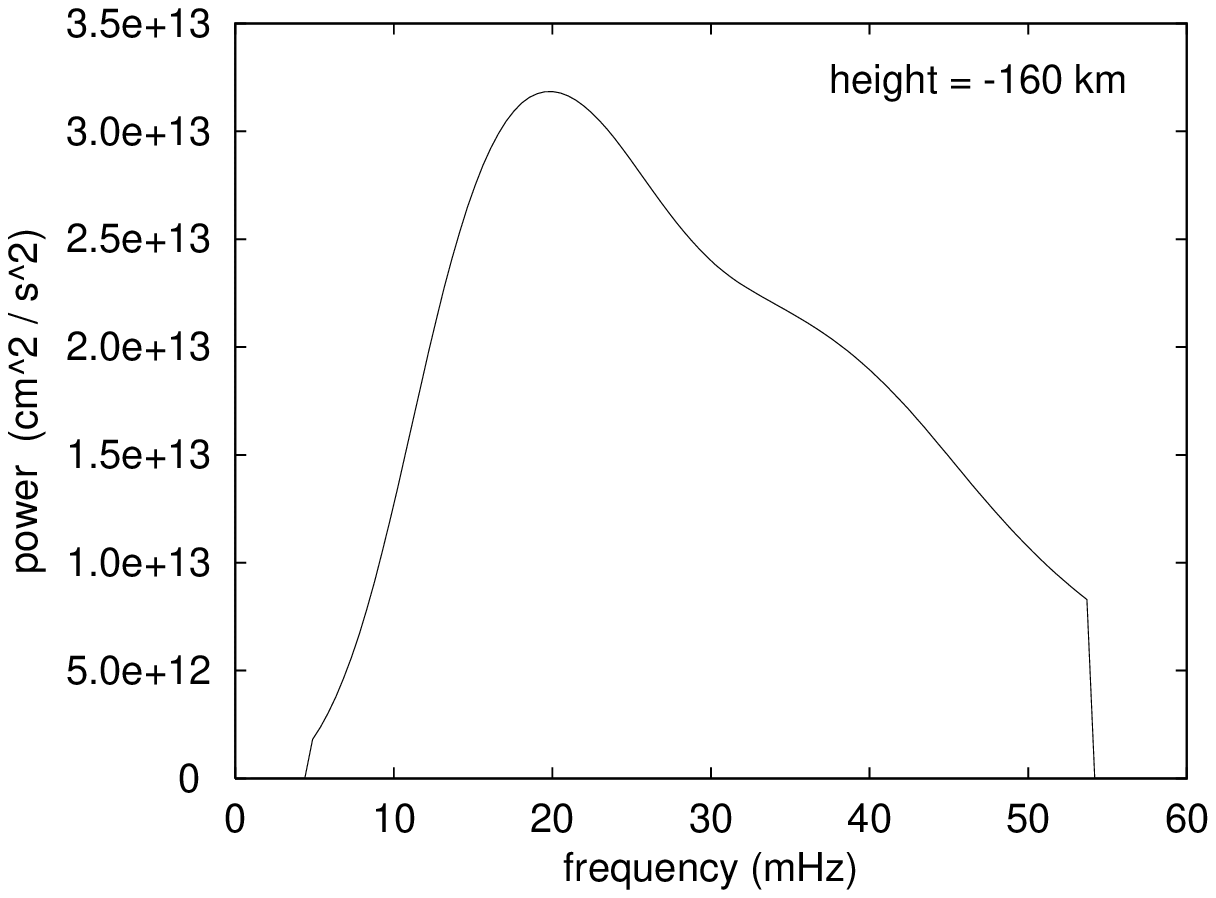,height=3.980cm,width=1.0 \textwidth}
\end{minipage}
\begin {minipage}[t]{0.33 \textwidth}
\psfig{figure=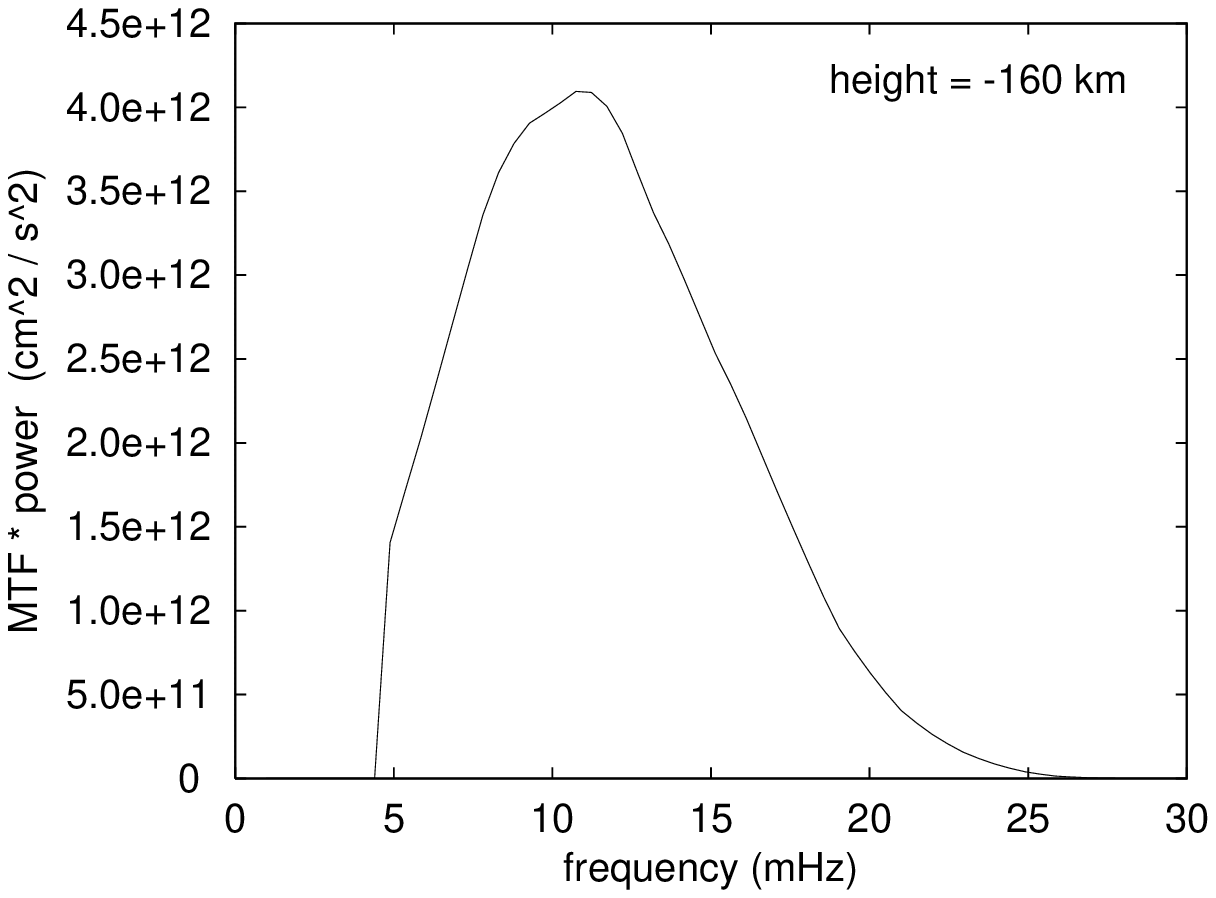,height=3.980cm,width=1.0 \textwidth}
\end{minipage}
\hfill
\begin {minipage}[t]{0.33 \textwidth}
\psfig{figure=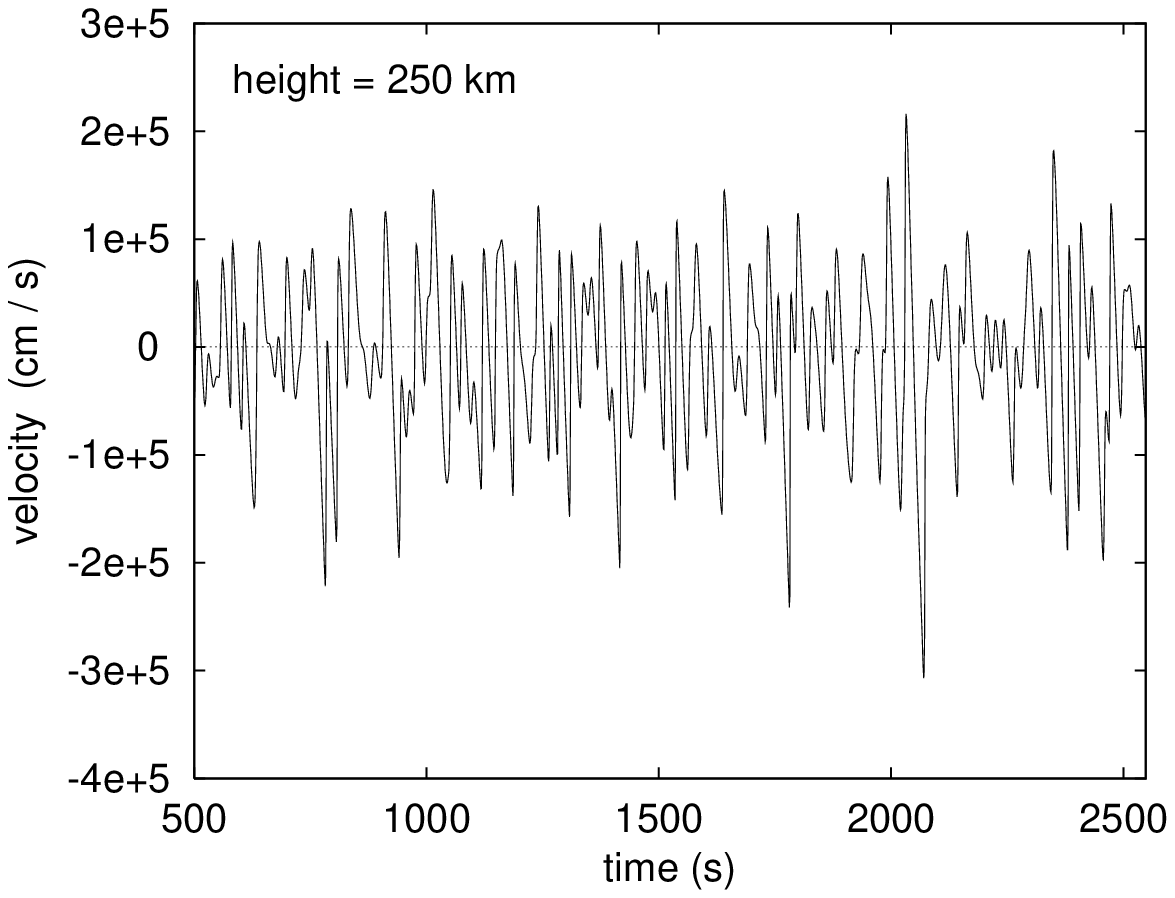,height=3.980cm,width=1.0 \textwidth}
\end{minipage}
\begin {minipage}[t]{0.33 \textwidth}
\psfig{figure=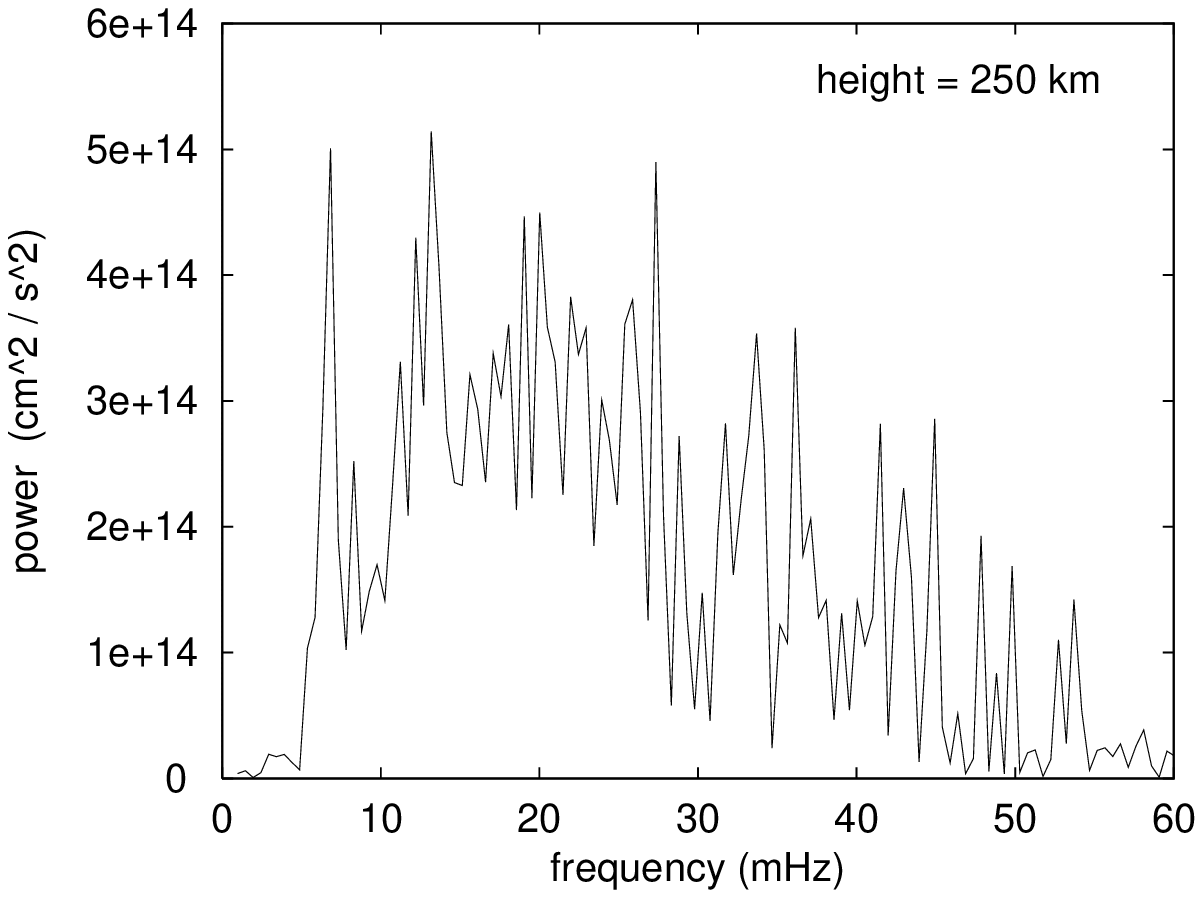,height=3.980cm,width=1.0 \textwidth}
\end{minipage}
\begin {minipage}[t]{0.33 \textwidth}
\psfig{figure=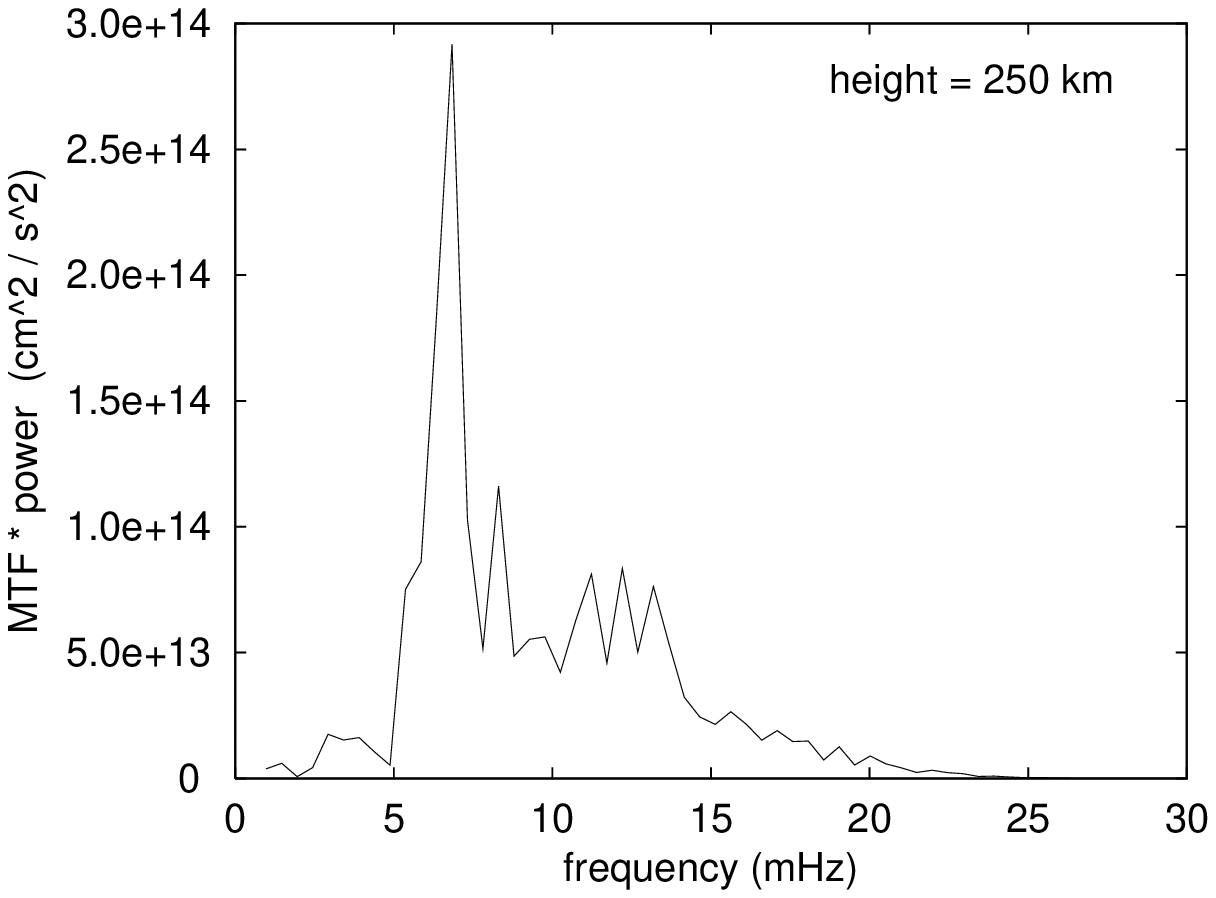,height=3.980cm,width=1.0 \textwidth}
\end{minipage}
\hfill
\begin {minipage}[t]{0.33 \textwidth}
\psfig{figure=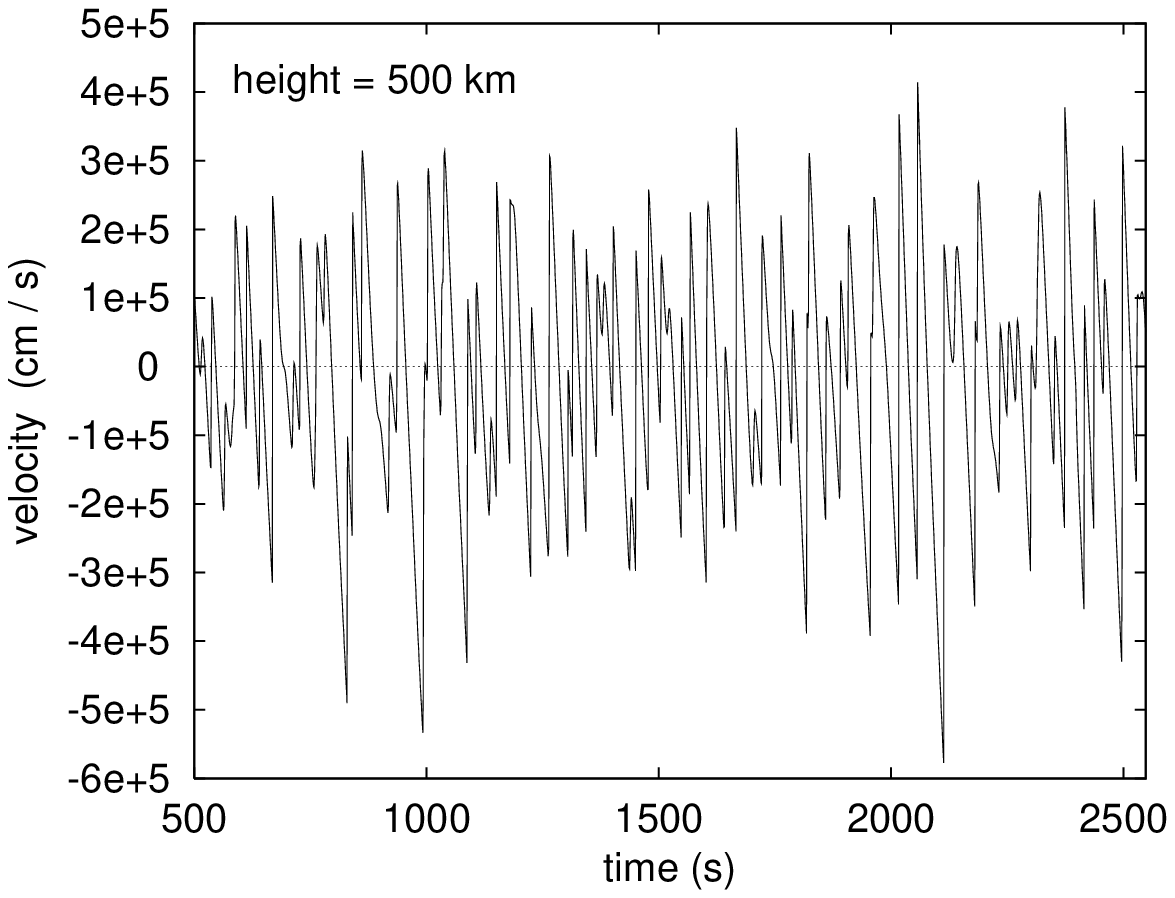,height=3.980cm,width=1.0 \textwidth}
\end{minipage}
\begin {minipage}[t]{0.33 \textwidth}
\psfig{figure=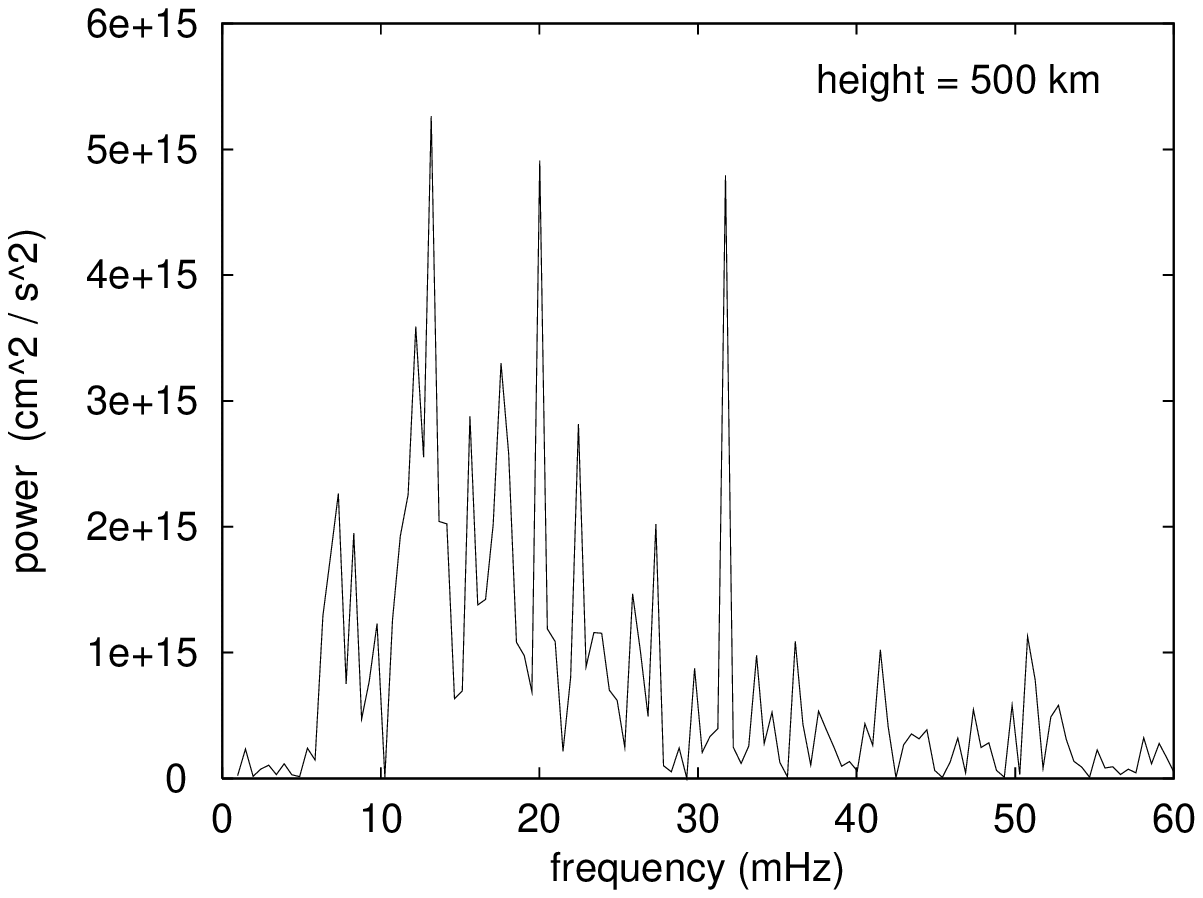,height=3.980cm,width=1.0 \textwidth}
\end{minipage}
\begin {minipage}[t]{0.33 \textwidth}
\psfig{figure=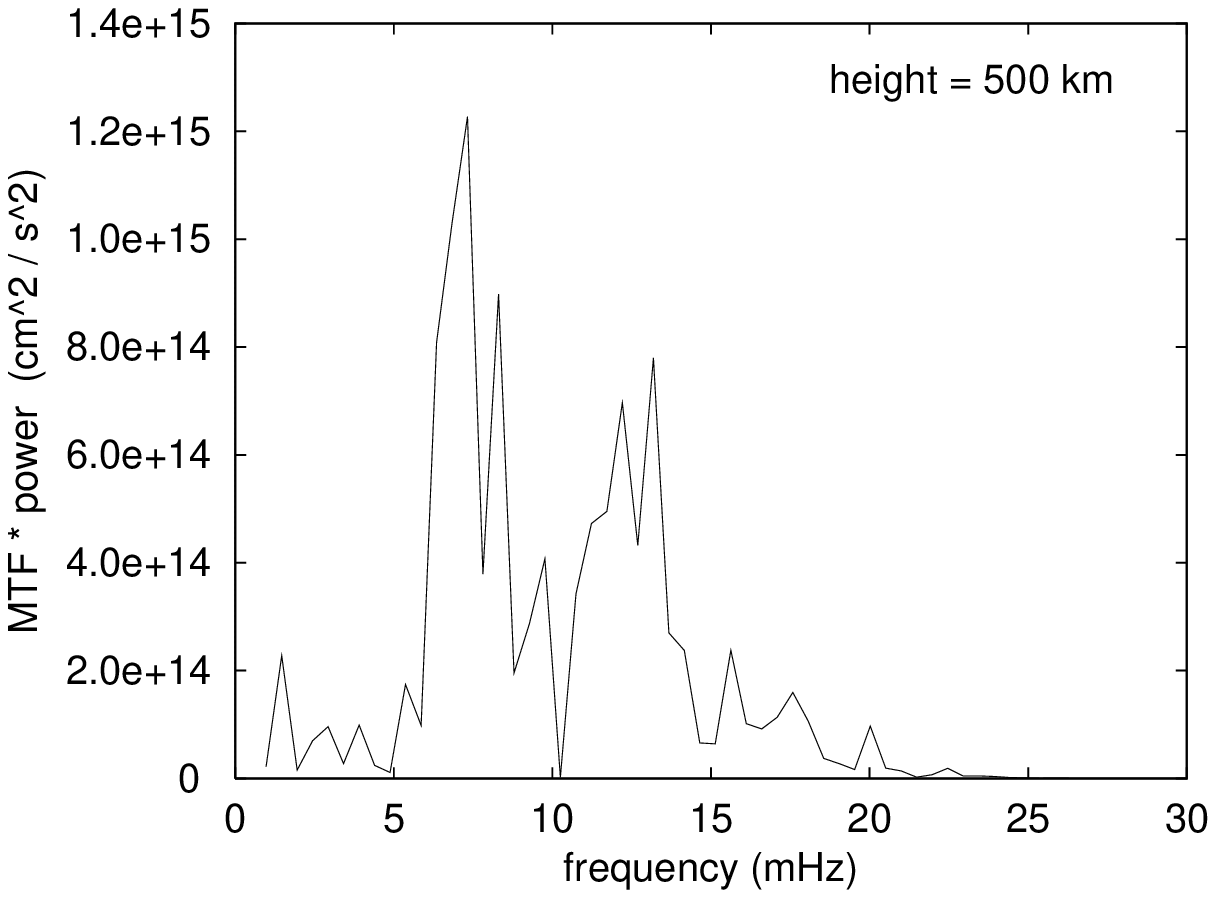,height=3.980cm,width=1.0 \textwidth}
\end{minipage}

\caption{
Velocities (left column), power spectra without (middle column) 
and power spectra with the applied modulation transfer function
at heights $z=- 160$, $250$ and $500$~km (from top to bottom).
The spectrum at $z = -160$~km is from a theoretical sound generation
calculation using the {\it extended Kolmogorov} (eKmG) turbulent energy 
spectrum in the solar convection zone. The acoustic spectrum has a 
total energy flux of $F_{\rm A}=1.97\cdot 10^8$~erg~cm$^{-2}$~s$^{-1}$.} 
\end {figure*} 

At this point we want to emphasize that because of the
modulation transfer function we do not consider simulations like
those of Carlsson \& Stein (1994, 1995) and Cheng \& Yi (1996)
to be very realistic, because they use a spectrum observed
from Earth as an input for their computations, leaving out the
largest part of the high frequency acoustic spectrum which
theoretically must be present at their lower boundary as
discussed above. Moreover in similar computations as those by
the mentioned authors we find strong emission cores in the IRT
lines of Ca II. These emission cores, which are also found for
the Vernazza et al.~(1981) solar models if microturbulence is
neglected or only a small amount of microturbulence is used,
are not observed. Inclusion of a large amount (see Vernazza et
al.) of microturbulence can supress these CA II IRT emission
cores. This microturbulence could possibly be the broad high
frequency band neglected by the above authors.

\section{Results} 
With our hydrodynamic code and using the spectra 
described above we have simulated
the propagation of acoustic waves from their 
generation height at $z = -160$~km up into the photosphere. 
Figures~3 and 4 show the resulting acoustic spectra at $z=-160$~km, 
$250$~km and $500$~km. The left-hand columns of these figures  
show the velocity fluctuations at the respective heights. It is 
seen that the velocities at any given height vary in a 
random way. The middle columns display the raw acoustic spectra 
while the right-hand columns show them with the modulation 
transfer function applied. Note that for the eKmG 
spectrum there is a much larger high-frequency contribution to 
the acoustic spectrum than for the rKmG spectrum. In addition, 
due to our choice of frequency spacing of the 101 partial waves 
it is seen that the acoustic spectra extend from 4~mHz to 54~mHz. 

Although the high-frequency filtering effect of the modulation 
transfer function cannot be observed at the generation height
($-160$~km) we have applied the transfer function to the input 
spectrum as well in order 
to show the magnitude of this effect. Note 
the difference in the frequency scales of the figures in the 
right-hand columns as compared to the middle columns. 
\begin{figure*}[t] 
\begin {minipage}[t]{0.33 \textwidth}
\psfig{figure=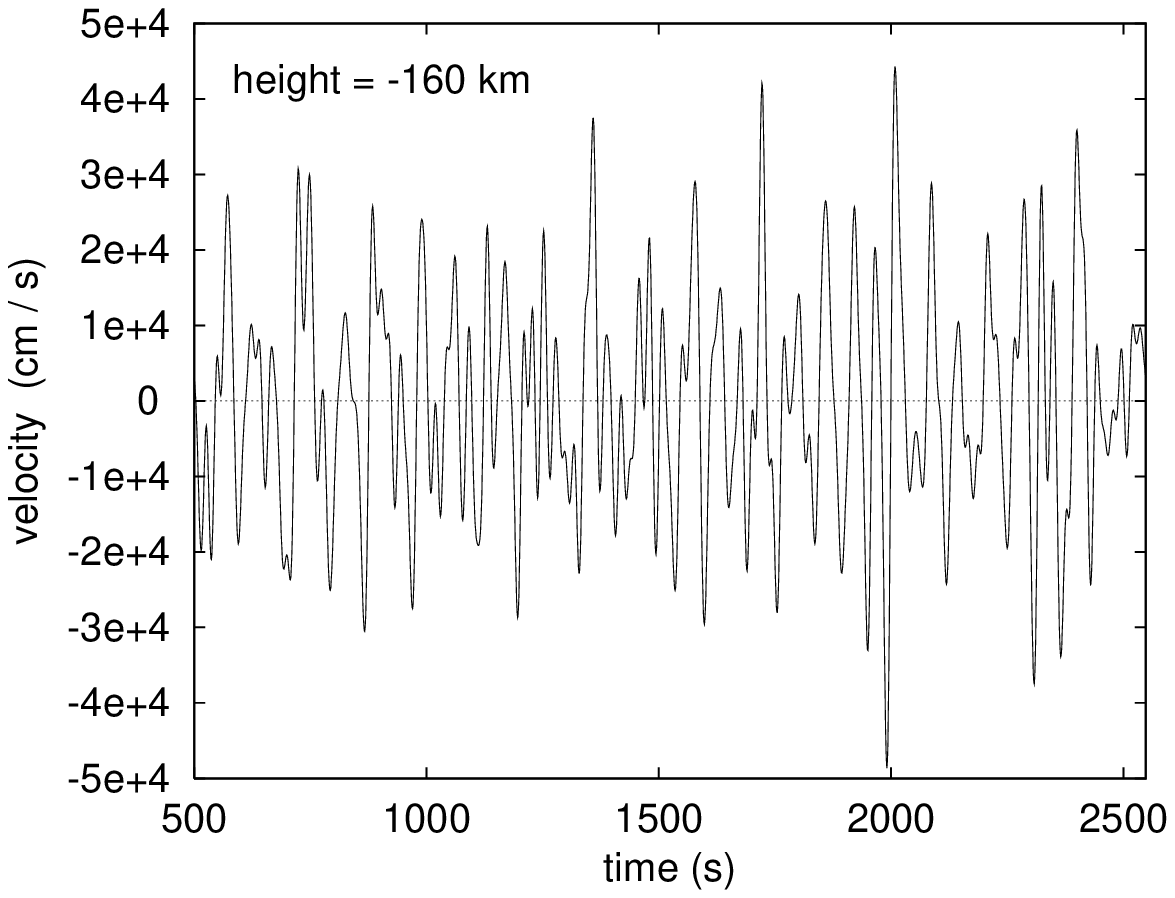,height=3.980cm,width=1.0 \textwidth}
\end{minipage}
\begin {minipage}[t]{0.33 \textwidth}
\psfig{figure=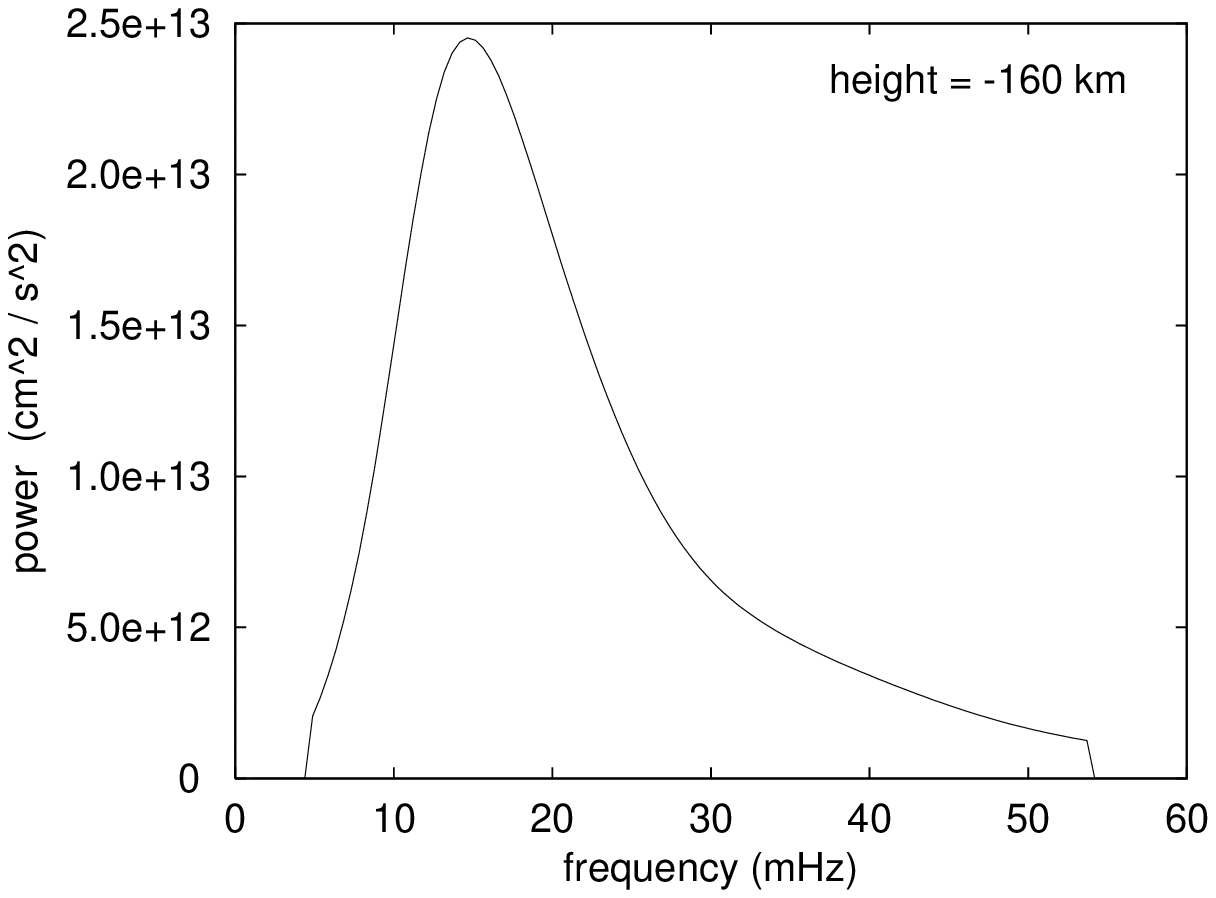,height=3.980cm,width=1.0 \textwidth}
\end{minipage}
\begin {minipage}[t]{0.33 \textwidth}
\psfig{figure=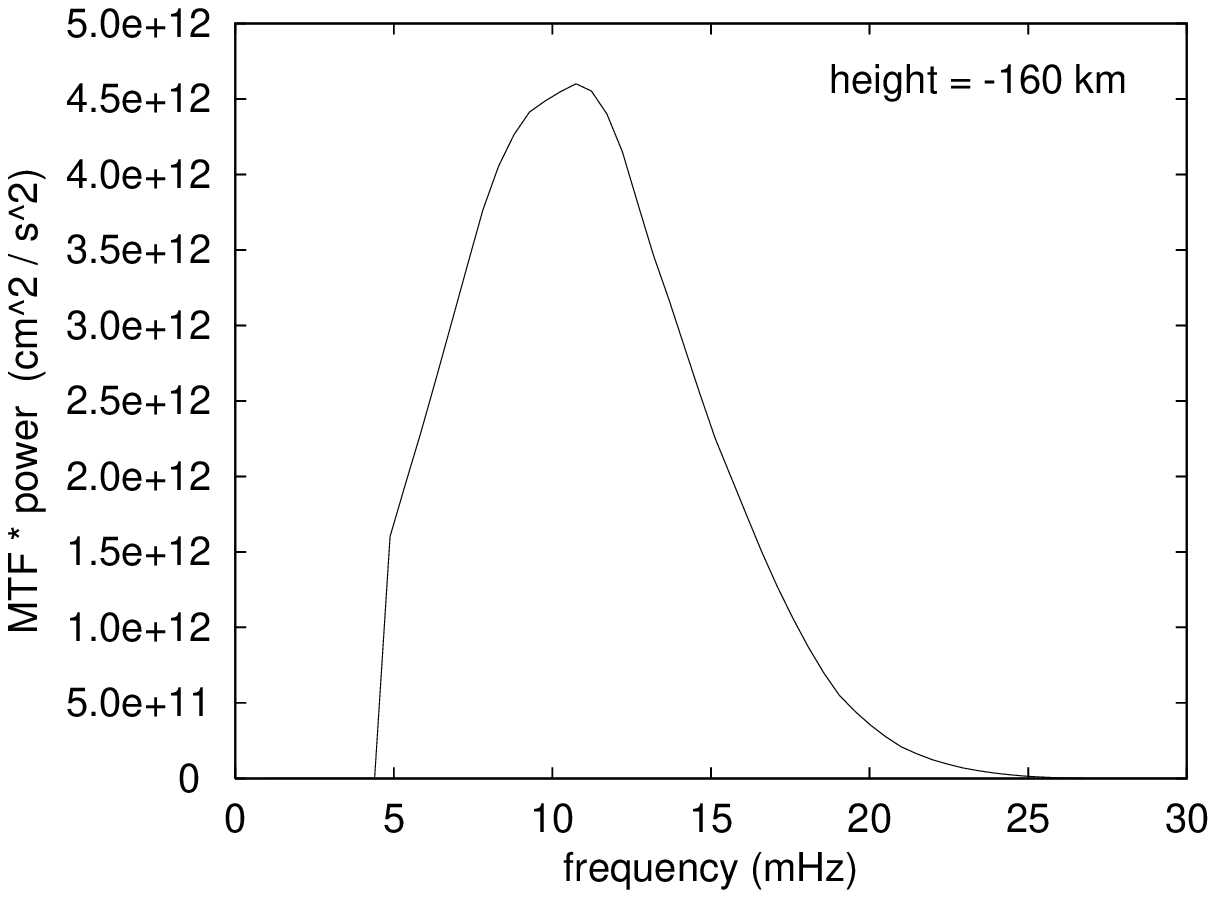,height=3.980cm,width=1.0 \textwidth}
\end{minipage}
\hfill
\begin {minipage}[t]{0.33 \textwidth}
\psfig{figure=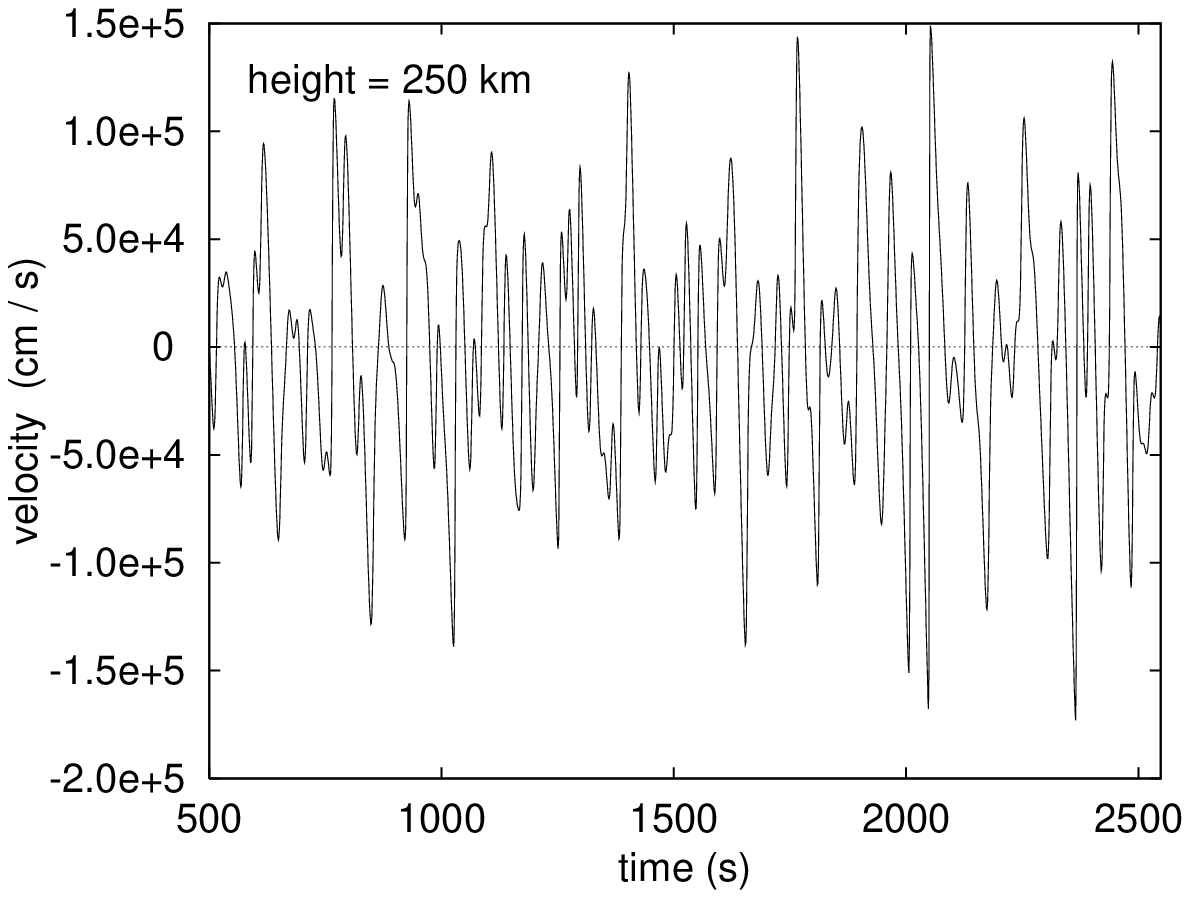,height=3.980cm,width=1.0 \textwidth}
\end{minipage}
\begin {minipage}[t]{0.33 \textwidth}
\psfig{figure=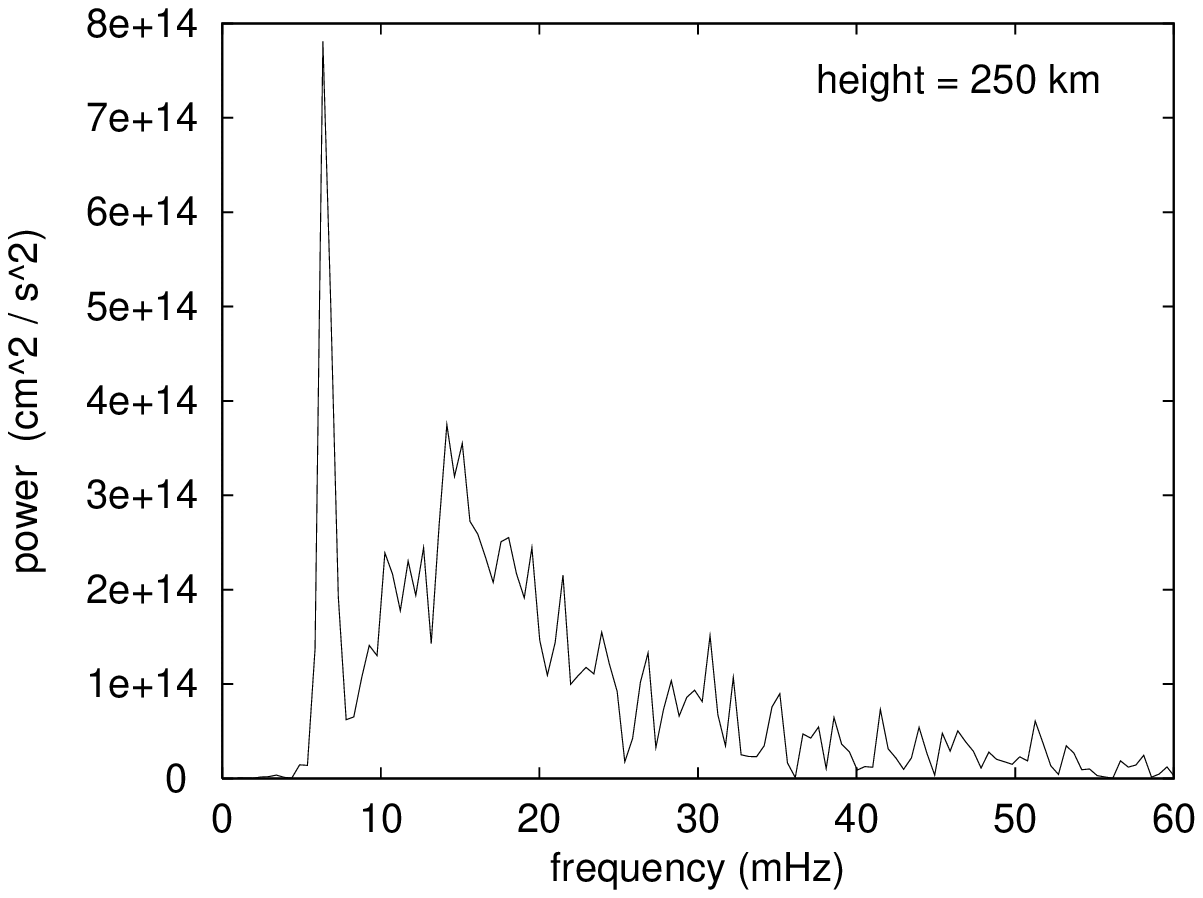,height=3.980cm,width=1.0 \textwidth}
\end{minipage}
\begin {minipage}[t]{0.33 \textwidth}
\psfig{figure=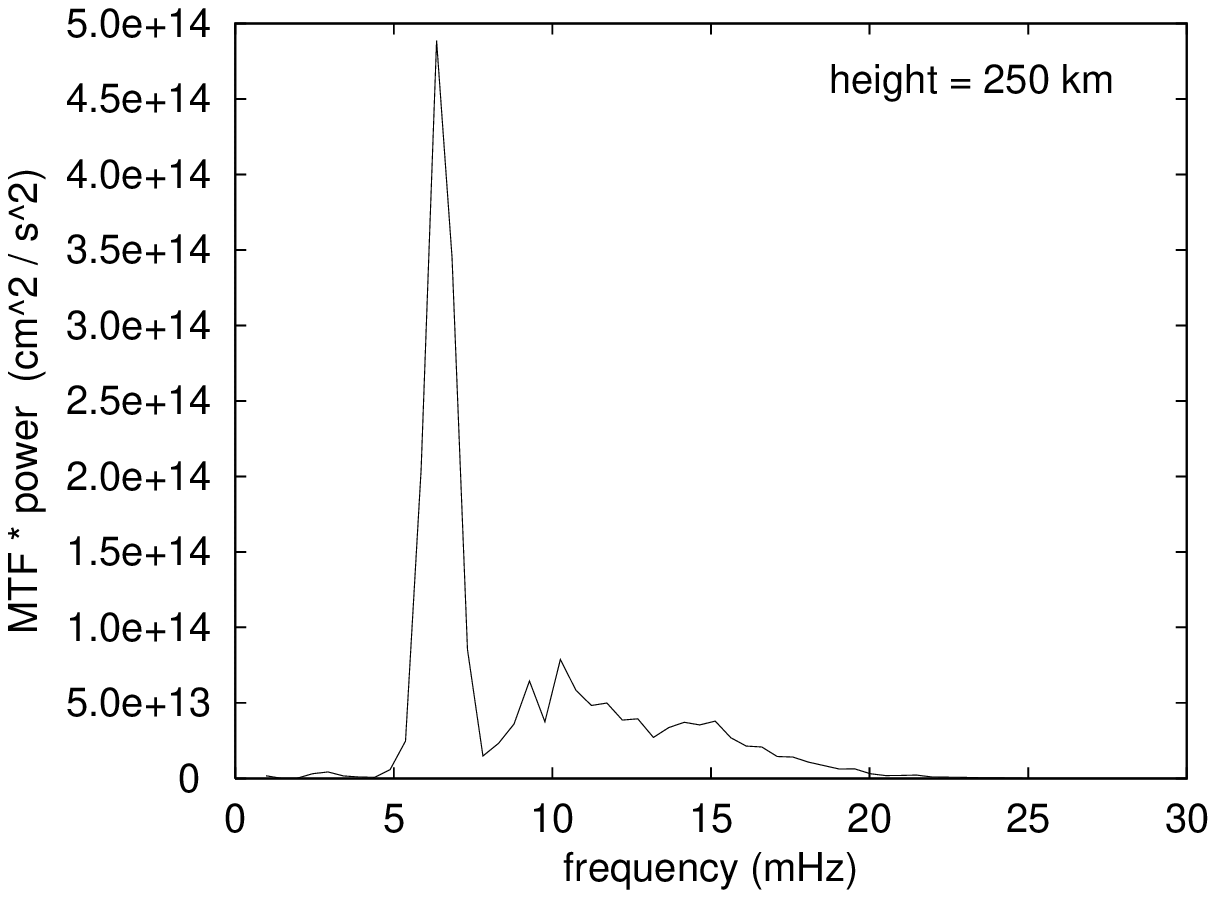,height=3.980cm,width=1.0 \textwidth}
\end{minipage}
\hfill
\begin {minipage}[t]{0.33 \textwidth}
\psfig{figure=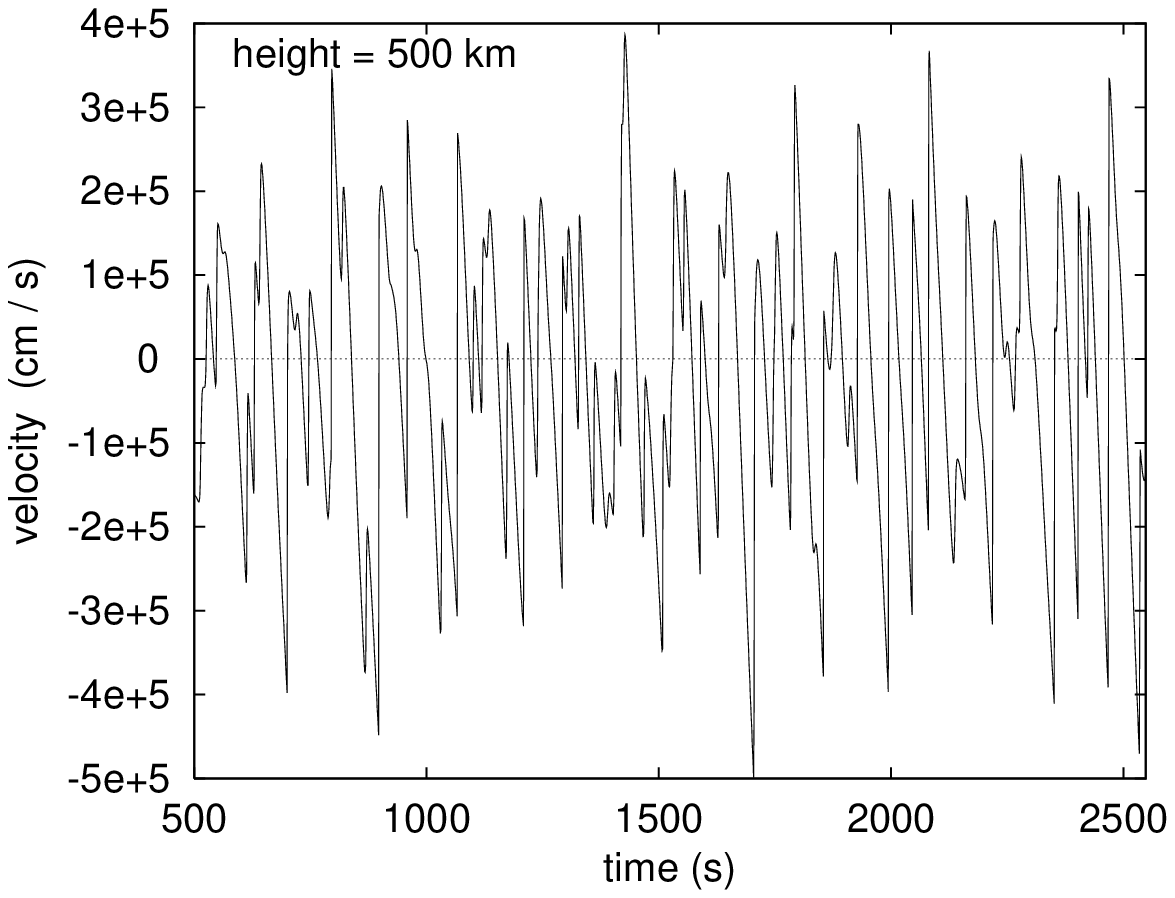,height=3.980cm,width=1.0 \textwidth}
\end{minipage}
\begin {minipage}[t]{0.33 \textwidth}
\psfig{figure=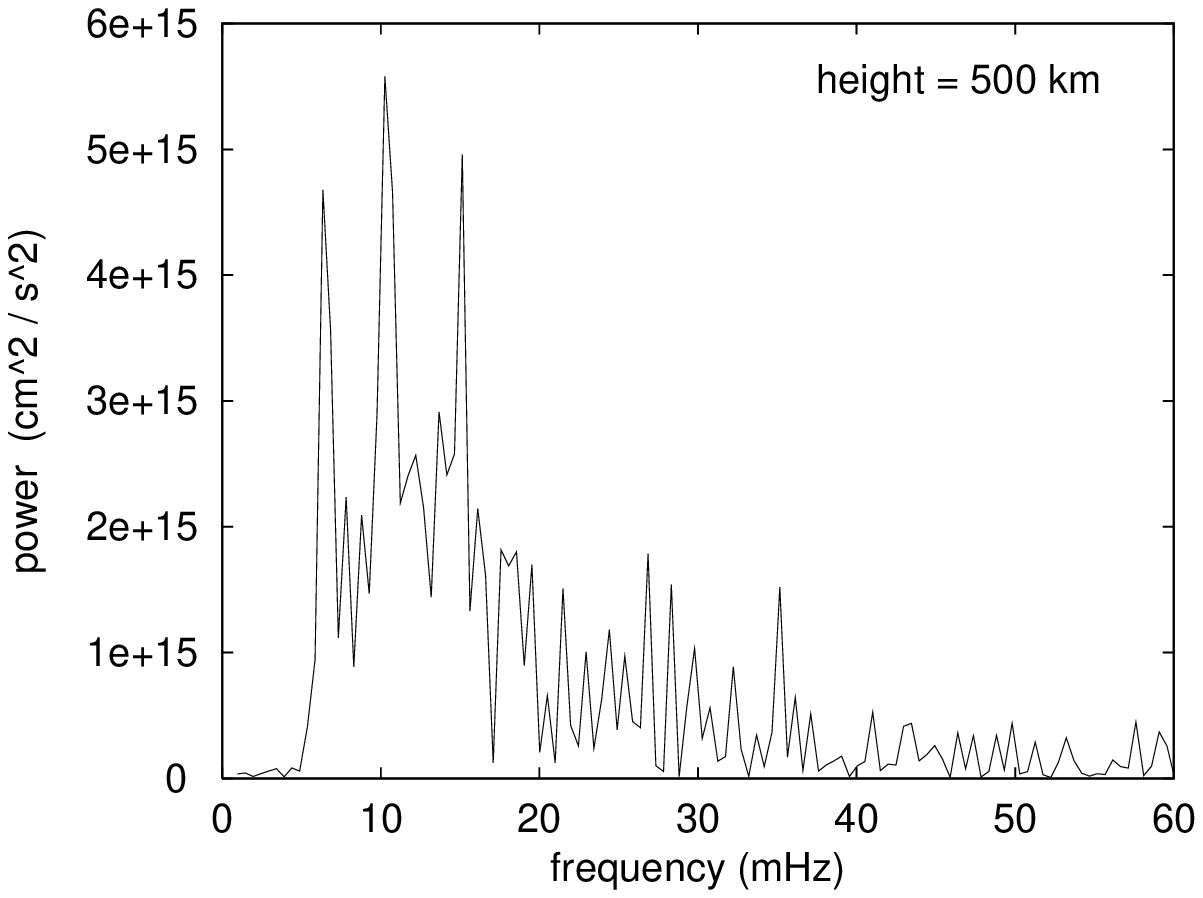,height=3.980cm,width=1.0 \textwidth}
\end{minipage}
\begin {minipage}[t]{0.33 \textwidth}
\psfig{figure=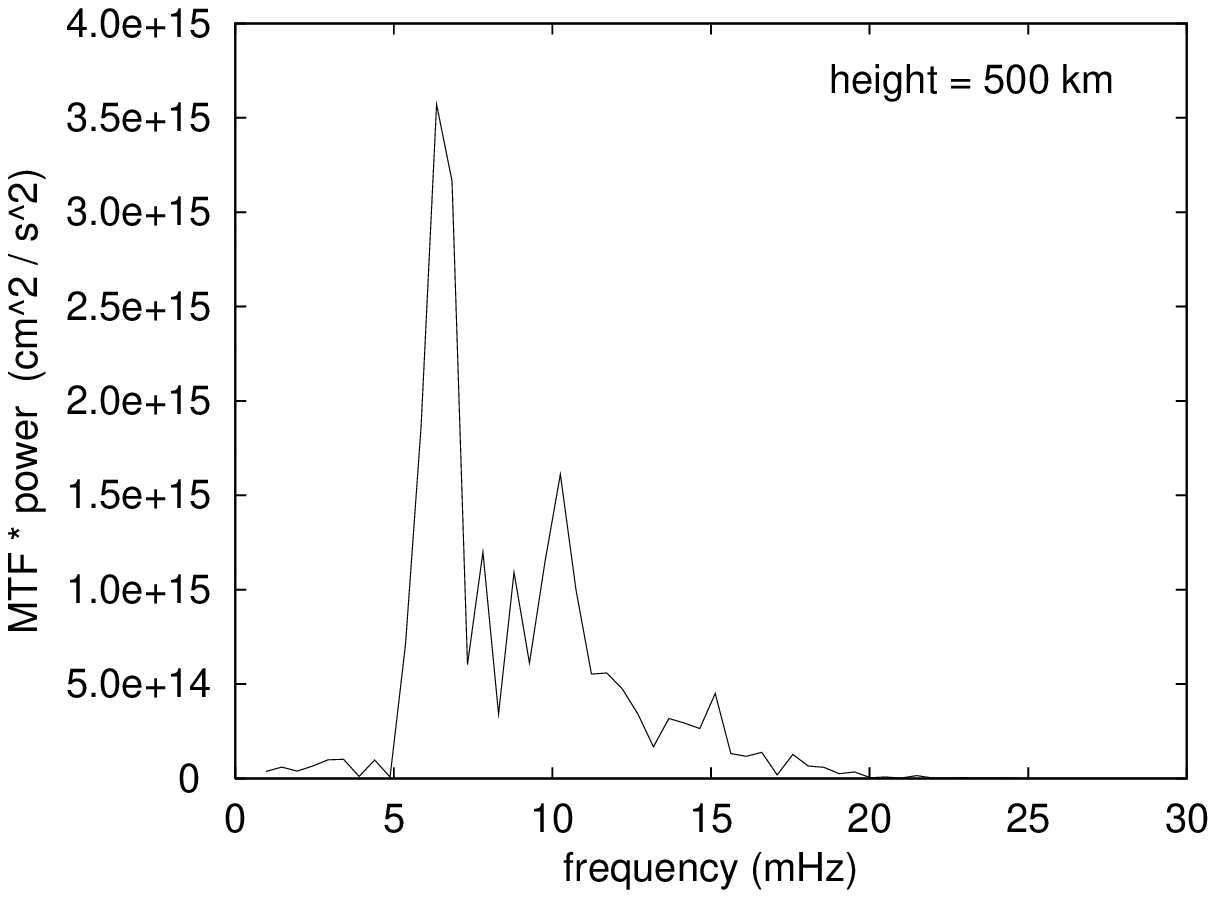,height=3.980cm,width=1.0 \textwidth}
\end{minipage}
\caption{
Velocities (left column), power spectra without (middle column) 
and power spectra with the applied modulation transfer function
at heights $z=- 160$, 250 and 500~km (from top to bottom).
The spectrum at $z = -160$~km is from a theoretical sound generation
calculation using the {\it raised Kolmogorov} (rKmG) turbulent energy 
spectrum in the solar convection zone. The acoustic spectrum has a 
total energy flux of $F_{\rm A}=8.75\cdot 10^7$~erg~cm$^{-2}$~s$^{-1}$.} 
\end {figure*} 
%

Comparing the spectra at heights of $250$~km and $500$~km with those 
at $-160$~km it is seen that the spectra 
progressively gain low-frequency components with height; 
this is best seen 
in the figures with the modulation transfer function applied. The 
spectra without the modulation transfer function 
retain a considerable contribution by 
high-frequency components and even 
largely their initial shape. But 
with increasing height, as already found in Paper IV, the 
initial spectral shape is progressively lost in favour of 
a spectrum with an increasingly 
dominant 3~min band.

A comparison of Figs.~1 and 3 shows rough agreement, but 
our simulations have more high-frequency power 
than the observations. But we should perhaps not overemphasize the 
differences between observation and theory in this exploratory 
calculation because of our rather simplistic approach where in 
the theoretical treatment we neglected radiative damping, 
the detailed evaluation of the modulation transfer function 
and seeing and instrumental effects, and where in the 
treatment of the 
observations we made several simplifying assumptions. The 
important result from the comparison of Figs.~3 and 4 with Fig.~1 
is that there is a dominant peak 
near 3~min in both the 
simulations and the observations and that this 
peak is not present in the initial acoustic spectra 
at the generation height of $-160$~km. This indicates that 
the generation of the 3~min waves in the simulations is 
a property of the atmosphere. 

It is remarkable that the low-frequency spectrum 
with a peak at 7~mHz derives from acoustic spectra peaked 
at high frequencies (of $\approx
14$ to 20~mHz). This result is at variance with 
the finding of Cheng \& Yi 
(1996) who claim that short-period waves play no role in the 
generation of the observed spectrum dominated by 
the 3~min peak, a signal they believe to be necessary
at site of the acoustic wave generation. 

As in Paper IV, we have monitored the heights of 
shock formation and of shock merging. In computations 
extending to a time of 5000~s we found that,
respectively, 190 and 150 
shocks were generated at heights concentrated between 200 to 
400~km and between 300 to 500~km for the eKmG and rKmG spectra. 
In addition, respectively, 80 and 60 shock mergings occurred at 
heights above 500 and 600~km for the eKmG and rKmG spectra. 
Temperature minima were generated at  350 to 
450~km in these calculations with temperatures reaching 
22000~K at 1500~km height in both cases. 

In our computations we also monitored the total acoustic flux as
function of height. We found that up to about $z=500$~km height
one essentially had flux conservation with the flux decreasing
by less than a factor of two, while above this height the flux
decreased rapidly due to the dissipation by fully developed
shocks. In this height interval of about 660~km, the density
decreased by more than two orders of magnitude, leading to a
large amplitude growth of the acoustic waves.

Another important result is that the acoustic fluxes computed on 
the basis of our simulations produce magnitudes in the 
power spectra ($3\cdot 10^{14}$ to $1.2\cdot 10^{15}$~cm$^2$/s$^2$
for the eKmG and $5\cdot 10^{14}$ to $3.5\cdot 
10^{15}$~cm$^2$/s$^2$ for the rKmG spectrum) that are in rough 
agreement with the magnitude of 
the observed power ($1\cdot 10^{15}$~cm$^2$/s$^2$) for
frequencies greater than 4~mHz. 
This is seen when comparing Figs.~3 and 4 
with Fig.~1 and indicates 
rough overall agreement between the theoretical concepts of
sound generation by turbulent convection and observation. 
This suggests that we might be able to
use the observations of acoustic velocity fluctuations to 
determine critical physical quantities like the maximum 
convective velocities, the depth of the maximal flow
velocity and the efficiency factors in 
the Lighthill-Stein theory of sound generation. The result 
moreover shows that a more refined theoretical simulation and a 
more detailed comparison with observation is urgently needed.

Theoretically the magnitude of the convective velocity appears 
to be not overly uncertain, as the comparison of numerical convection 
calculations with the mixing-length calculations by Steffen 
(1992) shows. He found that mixing-length computations provide 
maximum velocities of 2.0, 2.6 and 3.1~km/s for $\alpha$ =1, 2 and 
3, respectively, while his numerical convection calculations 
show maximum convective velocities of 2.8~km/s. A greater 
difference between the mixing-length results and his numerical 
convection simulations (which include radiation) is that in the 
former the convective velocities go to zero at the 
solar surface, while in the latter, due to the granular motion, 
the velocities remain finite into the photosphere, where they
decrease to a minimum value of 1.2~km/s. Such motions would
normally generate acoustic flux were it not for the $u^8$
dependence, which ignores all velocities except those close to 
the velocity maximum. Another significant difference of the 
mixing-length approach as compared with the numerical convection 
computations, as mentioned above, is that in the latter (Steffen
1992) the maximum convective velocity occurs about 100~km deeper.
This might allow the acoustic spectrum to show a stronger 3~min 
component because of the greater distance to the photosphere.
 
\section{Conclusions} 
Using adiabatic wave calculations we have followed the 
development of acoustic wave spectra from their  
generation height at $z = -160$~km in the solar convection 
zone, up into the photosphere, where we have compared them with 
observed spectra. 
  \newline
\noindent 1. We find that the acoustic spectra generated by the 
turbulence of the convection zone do not have a pronounced peak 
in the 3~min band. 
Instead, the acoustic spectra based on the extended 
(eKmG) and raised (rKmG) Kolmogorov spectra of turbulent 
convection and computed with convective velocities using the 
mixing-length theory with a mixing length parameter of $\alpha 
=2.0$ have single peaks at high frequency, at $20$ and 
$14$~mHz, respectively. 
  \newline
\noindent 2. The acoustic spectra in the simulation
acquire a predominant 3~min component 
after the propagation of the waves from the generation height at $-
160$~km into the photosphere at $250$ or $500$~km.
  \newline
\noindent 3. Up to these heights of $250$ to $500$~km the total
acoustic flux is essentially conserved, decreasing by less than
a factor of two. This shows that the shift from a high to a low
frequency spectrum is a true shift of the wave energy.
  \newline
\noindent 4. The theoretical spectra are in fair
agreement with the spectrum observed in an Fe I line at $z =
250$~km height after correction  using a modulation 
transfer function, which suppresses high-frequency spectral 
components that are unobservable in the Doppler effect 
due to the finite width of the line contribution function. 

\begin{acknowledgements}
We are grateful to the DFG, to NASA and NATO for support. We
are also very grateful to F.-L.~Deubner, S.~Steffens and
F.~Pijpers for comments on an earlier version of the manuscript.
\end{acknowledgements}


\begin{thebibliography}{qqq}
 
\bibitem{} Al N., 1996, Ph.~D.~thesis, Univ.~G\"ottingen, 
    Germany
 
\bibitem{} Buchsbaum G., 1987, Diplom thesis, Univ.~W\"urzburg, 
    Germany

\bibitem{} Carlsson M., Stein R.F., 1994. In: Carlsson M.~(ed.)
    Proc.~Oslo Mini-Workshop at Institut of Theoretical Astrophysics, 
    Chromospheric Dynamics, p.~47

\bibitem{} Carlsson M., Stein, R.F., 1995, ApJ 440, L29

\bibitem{} Cattaneo F., Brummell N.H., Toomre J., Malagoli A., 
    Hulburt N.E., 1991, ApJ 370, 282

\bibitem{} Cheng Q.-Q., Yi Z., 1996, A\&A 313, 971
 
\bibitem{} Deubner F.-L., 1991. In: Ulmschneider P., 
    Priest E.R., Rosner R.~(eds.) Mechanisms of Chromospheric and Coronal 
    Heating. Springer, Berlin, p.~6

\bibitem{} Deubner F.-L., Reichling M., Langhanki R.: 1988. 
  In: Christensen-Dalsgaard J., Frandsen S.~(eds.) IAU Symp.~123,
       Advances in Helio- and Asteroseis\-mology, p.~439
 
\bibitem{} Fleck B., Schmitz F., 1991, A\&A 250, 235

\bibitem{} H\"unsch M., Schr\"oder K.-P., 1997, A\&A 309, L51

\bibitem{} Kalkofen W., 1996, ApJ 468, L69

\bibitem{} Kalkofen W., Rossi P., Bodo G., Massaglia S., 1994, 
   A\&A 284, 976
 
\bibitem{} Kumar P., 1994, ApJ 428, 827

\bibitem{} Lites B.W., Rutten R.J., Kalkofen W., 1993, ApJ 414, 345

\bibitem{} Muller R., 1989. In: Rutten R.J., Severino G.~(eds.)
    Solar and Stellar Granulation. Kluwer, Dordrecht, p.~101
 
\bibitem{} Musielak Z.E., Rosner R., Stein R.F., Ulmschneider P., 
    1994, ApJ 423, 474 
  
\bibitem{} Nesis A., Hanslmeier A., Hammer R., et.~al., 1993, A\&A 279, 599

\bibitem{} Rossi P., Kalkofen W., Bodo G., Massaglia S., 1992. In:
    Giampapa M.S., Bookbinder J.A.~(eds.) Proc.~Seventh Cambridge Workshop,
    Cool Stars, Stellar Systems and the Sun. ASP Conf.~Series 26, p.~546

\bibitem{} Rutten R.J., 1995. In: Hoeksema J.T., Domingo V., Fleck B.,
     Battrick B.~(eds.),
     Proc.~4th SOHO Workshop Helioseismology. ESA SP-376, p.~151 
  
\bibitem{} Rutten R.J., 1996. In: Strassmeier K.G., Linsky J.L.~(eds.)
    IAU Symp.~ 176, Stellar Surface Structure. Kluwer, Dordrecht, p.~385
  
\bibitem{} Rutten R.J., Uitenbroek H., 1991, Solar Physics 134, 15
 
\bibitem{} Schmitz F., Fleck B., 1995, A\&A 301, 483

\bibitem{} Schr\"oder K.-P., Eggleton P.P., 1996, Reviews in Modern 
     Astronomy 9, 221

\bibitem{} Steffen M., 1992, Habilitation thesis, Univ.~Kiel, 
    Germany

\bibitem{} Steffens S., Deubner F.-L., Hofmann J., Fleck B., 1995, 
A\&A 302, 277
 
\bibitem{} Sutmann G., Ulmschneider P., 1995a, A\&A 294, 232 (Paper I)

\bibitem{} Sutmann G., Ulmschneider P., 1995b, A\&A 294, 241 (Paper II)
  
\bibitem{} Sutmann G., Musielak Z.E., Ulmschneider P., 1996, A\&A (submitted)
       (Paper III)

\bibitem{} Theurer J., Ulmschneider P., Cuntz M., 1997, A\&A 
(in press) (Paper IV) 

\bibitem{} Ulmschneider P., 1990. In: G. Wallerstein (ed.) 
       Proc. Sixth Cambridge Workshop, Cool Stars,
       Stellar Systems, and the Sun.
       ASP Conf.~Series 9, San Francisco, p.~3

\bibitem{} Ulmschneider P., Schmitz F., Kalkofen W., Bohn H.U., 
1978, A\&A 70, 487

\bibitem{} Ulmschneider P., Theurer J., Musielak Z.E., 1996, 
A\&A 315, 212

\bibitem{} Vernazza J.E., Avrett E.H., Loeser R., 1981, ApJS 45, 635

\bibitem{} von Uexk\"ull M., Kneer F., 1995, A\&A 294, 252

\end{thebibliography}
\end{document}